\documentclass[onecolumn,authoryear]{els-mrw} 

\usepackage{amsmath,amssymb,amsfonts,amsthm,makeidx,graphicx}
\usepackage{txfonts}
\usepackage{helvet}
\usepackage{xspace}
\usepackage{slashed}
\usepackage{bbold}
\usepackage{bbm}
\usepackage{empheq}
\usepackage{mathrsfs}
\usepackage{xcolor}
\usepackage{siunitx}
\usepackage{subcaption}

\usepackage{hyperref}

\newcommand{\id}{\mathbbm{1}}
\newcommand{\calA}{\mathcal{A}\xspace}
\newcommand{\calG}{\mathcal{G}\xspace}
\newcommand{\calH}{\mathcal{H}\xspace}
\newcommand{\calL}{\mathcal{L}\xspace}
\newcommand{\calM}{\mathcal{M}\xspace}
\newcommand{\calN}{\mathcal{N}\xspace}
\newcommand{\calO}{\mathcal{O}\xspace}

\newcommand{\Lamqcd}{\ensuremath{\Lambda_\text{QCD}}\xspace}

\newcommand{\SU}[1]{\ensuremath{\text{SU}(#1)}}

\newcommand{\chipt}{$\chi$PT\xspace}
\newcommand{\Nc}{\ensuremath{N_c}\xspace}

\newcommand{\Tr}{\text{Tr}}

\newcommand{\Lext}{\calL_\text{ext}}

\newcommand{\gCL}{\mathtt{g}_A}
\newcommand{\NN}{\ensuremath{N\!N}\xspace}

\newcommand{\MS}{\ensuremath{\widetilde{\text{MS}}}\xspace}

\newcommand{\oneS}{\ensuremath{{{}^{1}\!S_{\!0}}}\xspace}
\newcommand{\threeS}{\ensuremath{{{}^{3}\!S_{\!1}}}\xspace}

\begin{document}

\chapter{Symmetries of QCD and their relevance for low-energy nuclear physics}\label{chap:QCDSym}

\author[1]{Matthias R.~Schindler}%

\address[1]{\orgname{University of South Carolina}, \orgdiv{Department of Physics and Astronomy}, \orgaddress{Columbia, SC 29208}}


\maketitle

\begin{glossary}[Nomenclature]
\begin{tabular}{@{}lp{34pc}@{}}
EFT & Effective field theory\\
LEC & Low-energy coefficient\\
LO & Leading order \\
NLO & Next-to-leading order \\
QCD & Quantum chromodynamics\\
QED & Quantum electrodynamics\\
\chipt & Chiral pertrubation theory
\end{tabular}
\end{glossary}

\begin{abstract}
    QCD, the theory of the strong interactions, is formulated in terms of quarks and gluons, while low-energy nuclear physics deals with hadrons such as protons, neutrons, and pions. 
    Symmetries establish a systematic connection between these two descriptions of strongly-interacting systems.
    The objective of this article is to review the symmetries of QCD and to explain how they  constrain hadronic interactions.
    Chiral symmetry, which emerges in QCD in the limit of massless quarks, is of particular importance for low-energy nuclear physics.
    Together with its explicit and spontaneous breaking, chiral symmetry provides the basis for chiral perturbation theory, the effective field theory describing pions and nucleons at low energies.
\end{abstract}

\begin{keywords}
 	QCD\sep Nuclear physics\sep  Symmetries\sep Effective field theory\sep Chiral perturbation theory
\end{keywords}

\section*{Objectives}
\begin{itemize}
	\item Identify the exact and approximate symmetries of QCD
    \item Analyze limits in which approximate symmetries become exact
    \item Explain how these symmetries constrain interactions between hadrons at low energies
\end{itemize}

\section{Introduction}
\label{sec:intro}

Low-energy nuclear physics studies how protons and neutrons interact and how  the forces between them lead to the formation of nuclei and to nuclear reactions.
In the following, low energy refers to energies well below \SI{1}{GeV}, which is approximately equivalent to the mass of the proton and neutron.\footnote{We use units in which $\hbar = c = 1$, so that energies and masses have the same dimension.}
The interactions between protons and neutrons are primarily governed by the strong interactions: 
Electromagnetic effects (e.g., the Coulomb repulsion between protons) and weak interactions are much smaller than the strong interactions at typical nuclear scales. While Coulomb effects can play an important role in nuclear structure, and weak interactions are responsible for processes such as beta decay, the nuclear states involved in these processes are predominantly determined by the strong interactions.

In our current understanding, the fundamental theory of the strong interactions is quantum chromodynamics (QCD) \citep{Fritzsch:1973pi,Gross:1973id,Weinberg:1973un}; for a recent comprehensive overview of QCD see \citep{Gross:2022hyw}.
Thus QCD should play a central role in low-energy nuclear physics.
However, a direct description of nuclear interactions in terms of QCD is highly nontrivial.
At low energies, the coupling of QCD is large, which means that, unlike e.g.~in quantum electrodynamics (QED), perturbation theory in the QCD coupling is not a useful approach.
In addition, QCD is formulated in terms of quarks (the matter fields) and gluons (the gauge bosons or force carriers), but these are never observed as free particles, due to the confinement property of QCD.  
The observed strongly-interacting particles, called hadrons, are composite many-body systems of quarks and gluons with highly complex internal structure. 

Even if possible, a description of low-energy nuclear physics in terms of quarks and gluons might not be the most efficient approach. 
In classical mechanics, we can describe the collision of two macroscopic objects without having to keep track of every quark, gluon, and electron in each object and the interactions between them.
Similarly, nuclear physics is described by effective theories of the strong interactions, in which hadrons such as protons, neutrons, pions, etc., rather than the unobserved quarks and gluons, serve as the relevant degrees of freedom.
The interactions between them are manifestations of the underlying QCD interactions between quarks and gluons, and one of the main goals of nuclear physics is to establish a connection between QCD and nuclear interactions.
Lattice QCD is a numerical approach to perform QCD calculations, and its application to nuclear physics provides important contributions to our understanding of how QCD and nuclear physics are connected \citep{EncycLattice,Beane:2010em,Lin:2015dga,Drischler:2019xuo}.
Another approach, the one we will follow here,  to establish a connection between QCD and nuclear physics is through the use of symmetries.

\begin{figure}[ht]
\centering
\begin{subfigure}{0.4\textwidth}
\centering
    \includegraphics[width=.2\textwidth]{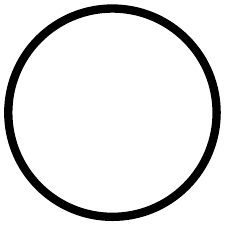}
    \caption{}\label{fig:circ}
\end{subfigure}
\begin{subfigure}{0.4\textwidth}
\centering
    \includegraphics[width=.2\textwidth]{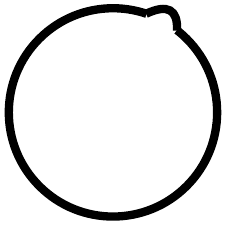}
    \caption{}\label{fig:bump}
\end{subfigure}
\caption{The object in (a) is invariant under rotations, while the one in (b) is not.}
\end{figure}

What do we mean by a symmetry?
In general, a symmetry corresponds to the invariance of an object when a transformation is applied.
Consider the circle of Fig.~\ref{fig:circ} as an example.
If we rotate it by any arbitrary angle within the plane about an axis through its center, it will still look the same -- the circle is invariant under rotations about the axis through its center.
The circle with the small bump in Fig.~\ref{fig:bump} on the other hand is not rotationally invariant.
After rotations by any angle other than multiples of $2\pi$ we will see a difference -- the bump will be in a different location. 
In other words, the bump has broken rotational invariance.
This type of breaking is called \emph{explicit} symmetry breaking, since it can be recognized by simply inspecting the object itself.
However, despite the bump, the original circle is still recognizable.
As long as the bump is very small compared to the circle itself, we can think of rotational symmetry as an \emph{approximate} symmetry and can try to include the bump as a small perturbation on top of the rotationally-invariant circle.
How well this approach works depends on how large the bump is relative to the circle. 
The smaller it gets, the better the approximate description of the figure as a circle.
This example contains several concepts we will encounter repeatedly: 
A symmetry is the invariance of an object under a transformation. 
An object might not be symmetric under some transformation, but a symmetry can emerge if a certain feature of the object is tuned; and as long as that feature is small, it might be possible to treat it as a perturbation.

In the following discussion, the objects are the Lagrangians of QCD and those of effective theories, as these provide the fundamental and effective theoretical descriptions of the strong interactions.
We will consider two general types of transformations: 
The first type  are changes of the spacetime coordinates $x^\mu = (t,\vec{x})$. 
These include, e.g., parity transformations under which $\vec{x} \mapsto -\vec{x}$ and time reversal $t\mapsto -t$ (the symbol $\mapsto$ is used to indicate a transformation). 
The second type are transformations applied to the fields that appear in the Lagrangian, $\phi(x) \mapsto \phi'(x)$, where $\phi(x)$ denotes a generic field. 
For example, fields are multiplied by a phase factor, or vectors of fields are multiplied by a matrix.
If the Lagrangian obtained after the transformation describes the same physics as the original Lagrangian, we say that the theory described by this Lagrangian has a symmetry.
In terms of the Lagrangian, this means that the Lagrangian is either invariant, i.e., its form does not change, or it at most changes by a total derivative,
\begin{align}
    \calL(\phi,\partial_\mu\phi) \mapsto \calL(\phi',\partial_\mu\phi') + \partial_\mu J^\mu.
\end{align}
The change by a total derivative is allowed because the physics is governed by the action $S=\int d^4 x \, \calL $.
With suitable boundary conditions, the total derivative in the Lagrangian corresponds to an irrelevant surface term.

How do symmetries help in establishing a connection between the underlying theory of QCD and effective descriptions of the strong interactions?
Because the degrees of freedom used in QCD and in the effective theories are different, it is not always obvious how a transformation at the quark level corresponds to transformation of the effective fields.
Nonetheless, we expect that a symmetry of QCD should also be a symmetry of any effective description.
Imposing the symmetries of QCD onto the effective interaction then implies constraints on the form the effective Lagrangian.
For example, assume that QCD is invariant under a transformation that corresponds to each hadronic field transforming as $\phi_i(x) \mapsto -\phi_i(x)$.
The invariance of QCD then imposes the constraint that the effective Lagrangian can only contain terms with an \emph{even} number of fields.
The more symmetries of QCD we can identify and incorporate in our effective descriptions, the more closely these effective descriptions should reproduce the results one would obtain from QCD.

In the context of quantum field theory, this approach is known as the effective field theory (EFT) framework, to be briefly discussed in Sec.~\ref{sec:EFT}. 
EFTs have become a standard tool in particle and nuclear physics.
They are discussed in more detail in the contribution \citep{EncycEFT}; for textbook treatments, see, e.g., \citep{Petrov:2016azi,Burgess:2020tbq,Meissner:2022cbi,Brauner:2024juy}.
The EFT of QCD that describes the interactions of hadrons at low energies is chiral perturbation theory (\chipt) \citep{Weinberg:1978kz,Gasser:1983yg,Gasser:1984gg}, which is discussed in Secs.~\ref{sec:mesonchpt} and \ref{sec:bchpt}.
For pedagogical introductions and review articles, see., e.g., \citep{Leutwyler:1993iq,Bernard:1995dp,Bernard:2006gx,Scherer:2012xha,Meissner:2022cbi,Meissner:2024ona} and references therein.

Many of the symmetries discussed here were known from observations before the development of QCD in 1973 \citep{Fritzsch:1973pi,Gross:1973id,Weinberg:1973un}.
For example, isospin (see Sec.~\ref{sec:isospin}) was first proposed  in 1932 as a symmetry involving protons and neutrons \citep{Heisenberg:iso}, but is now understood in terms of rotations in the space of $u$ and $d$ quarks (isospin space).
Similarly, chiral symmetry (see Sec.~\ref{sec:SymMassless}) was first discussed in terms of hadronic theories starting in the early 1960s  (e.g., \citep{Nambu:1960xd,Nambu:1961tp,Gell-Mann:1962yej}), but can now be shown to be an approximate symmetry of QCD.
QCD has provided a unifying theoretical foundation for these approximate symmetries and their possible breaking: they are properties of the quantum field theory underlying the strong interactions.

\section{QCD Lagrangian}
\label{sec:QCDLag}

As discussed above, the symmetries of QCD constrain the form of the effective interactions that are used in low-energy nuclear physics.
To identify these symmetries, we begin by analyzing the QCD Lagrangian.
Quarks are spin-1/2 particles that come in six so-called flavors:  up ($u$), down ($d$), strange ($s$), charm ($c$), bottom ($b$), and top ($t$).
In addition, each quark flavor carries a quantum number called color, which can take three values. 
The QCD Lagrangian can be obtained from the Lagrangian of free (noninteracting) quarks by imposing a symmetry: invariance under \emph{local} $\SU{3}_c$ transformations that act on the color degree of freedom of the quarks.\footnote{The subscript $c$ stands for ``color''  and is used to distinguish this group from other \SU{3} transformations  we will encounter.} 
This also necessitates the introduction of gluons, the gauge bosons of QCD, and determines the form of the QCD interactions.
The resulting QCD Lagrangian density (Lagrangian densities are simply referred to as Lagrangians in the following) is given by\footnote{The notation of Eq.~\eqref{eq:fullQCDLag} is very compact. In addition to carrying the flavor index $f$, the quark fields are four-component Dirac spinors with a spinor index $\alpha=1,\ldots,4$ that also carry a color index $A=1,\ldots,3$. Since the following discussion focuses on symmetries in flavor space, neither the spinor nor the color indices are shown to avoid notational clutter. We use Feynman's ``slash'' notation $\slashed{a}=a^\mu \gamma_\mu$, where $a_\mu$ is a four-vector and the $\gamma_\mu$ denote the Dirac matrices. The field strength tensor $\calG_{\mu\nu}$ is a matrix in color space and the subscript $c$ on the trace symbol denotes that it is taken in color space. The Lagrangian with all indices explicitly shown can be found, e.g., in \cite{Scherer:2012xha}.}
\begin{align}
\label{eq:fullQCDLag}
    \calL_\text{QCD} = \sum_{f=u,d,s,c,b,t} \bar{q}_f \left( i\slashed{D}-m_f \right) q_f -\frac{1}{2}\text{Tr}_c\left(\calG_{\mu\nu}\calG^{\mu\nu}\right),
\end{align}
where $m_f$ denotes the mass of a quark of flavor $f$.
The covariant derivative of the quark fields governs quark-gluon interactions.
It is given by 
\begin{align}
    D_\mu q_f = (\partial_\mu +i g_s \calA_\mu)q_f ,
\end{align}
with $g_s$ the strong $\SU{3}_c$ coupling and $\calA_\mu$ the gluon fields.
The gluon field strength tensor $\calG_{\mu\nu}$ in Eq.~\eqref{eq:fullQCDLag} is given by
\begin{align}
    \calG_{\mu\nu}=\partial_\mu \calA_\nu-\partial_\nu \calA_\mu +i g_s[\calA_\mu,\calA_\nu].
\end{align}

In the following, our focus will be on low-energy nuclear physics. 
By this we mean that our external states will consist of particles like protons, neutrons, pions, kaons, etc.,  with masses below or on the order of  \SI{1}{GeV}. 
The energies of these particles will also be restricted to remain below \SI{1}{GeV}.
This simplifies the situation.
The masses of the six different quark flavors vary widely, by more than four orders of magnitude, see Table~\ref{tab:qmasses}.
They fall into two groups: the light quarks $u$, $d$, and $s$ with masses well below \SI{1}{GeV}, and the heavy quarks with masses at and well above that scale.
\begin{table}[t]
\TBL{\caption{Approximate quark masses; see \citep{PDG} for details.}\label{tab:qmasses}}
{\begin{tabular*}{\textwidth}{@{\extracolsep{\fill}}@{}lllllll@{}}
\toprule
Quark flavor $f$ & $u$ & $d$ & $s$ & $c$ & $b$ & $t$\\
\colrule
Quark mass $m_f$ [GeV] & 0.002 & 0.005 & 0.093 & 1.27 & 4.18 & 173 \\
\botrule
\end{tabular*}}{%
}%
\end{table}
Thus, for our purposes of low-energy nuclear physics the heavy $c$, $b$, and $t$ quarks do not play a dynamical role, and we only consider light quarks explicitly.
(Technically, the heavy quarks are integrated out in the Feynman path integral sense; see, e.g., ~\citep{Donoghue:2022wrw}, and their effects are encoded into the parameters of the low-energy theories.) 
In the restriction to only light quark flavors, we will distinguish two cases: one in which the sum over quark flavors in Eq.~\eqref{eq:fullQCDLag} only includes $u$ and $d$ quarks; the other in which the $s$ quark is also considered.
To help identify additional symmetries of QCD, it is convenient to rewrite the Lagrangian further.
Depending on the case, the light quark fields can be collected into a two- or three-component vector $q$ (color and Lorentz indices are again not shown in the following), 
\begin{align}
	q=\begin{pmatrix} q_u \\ q_d \end{pmatrix} \quad \text{or} \quad q=\begin{pmatrix} q_u \\ q_d \\q_s \end{pmatrix}.
\end{align}
The QCD Lagrangian with only light quark flavors then takes the very compact form
\begin{align}
\label{eq:QCDLagLight}
    \calL_\text{QCD}^l &= \bar{q} \left( i\slashed{D}-\calM \right) q -\frac{1}{2}\text{Tr}_c\left(\calG_{\mu\nu}\calG^{\mu\nu}\right),
\end{align}
where $\calM$ denotes the quark mass matrix,
\begin{align}
\label{eq:qMassMat}
	\calM = \begin{pmatrix} m_u & 0 \\
	0 & m_d \end{pmatrix} \quad \text{or} \quad\calM = \begin{pmatrix} m_u & 0 & 0 \\
	0 & m_d & 0 \\
	0 & 0 & m_s
	\end{pmatrix} .
\end{align}
The Lagrangian $\calL_\text{QCD}^l$ of Eq.~\eqref{eq:QCDLagLight} will serve as the basis of most of the following discussion.

While $\SU{3}_c$ color gauge invariance is the defining symmetry of QCD, it does not play an explicit role in low-energy nuclear physics. 
The relevant degrees of freedom at low energies are hadrons like protons, neutrons, and pions, and color gauge symmetry underlies the formation of these particles. 
But because each hadron is a color-neutral object, $\SU{3}_c$ gauge transformations act trivially at the hadronic level and do not provide any additional direct constraints on their effective interactions.

However, as we will discuss in Sec.~\ref{sec:largeN}, one can gain additional information from color gauge symmetry by considering a generalization of QCD in which the number of colors is changed from 3 in the real world to a large number \Nc. This also means that the gauge symmetry of QCD changes from $\SU{3}_c$ to \SU{\Nc}. 
In the large-\Nc limit, new constraints on the effective hadronic interactions arise.

\section{Discrete symmetries}
\label{sec:discrete}

As a Lorentz-invariant local quantum field theory, QCD is invariant under $CPT$ transformations, i.e., the combined discrete transformations of charge conjugation ($C$), parity ($P$), and time reversal ($T$) \citep{Streater:1989vi}. 
Charge conjugation $C$ interchanges particles with antiparticles, while $P$ and $T$ are spacetime transformations with
\begin{align}
    P: \quad & (t,\vec{x}) \overset{P}{\longmapsto} (t,-\vec{x}),\\
    T: \quad & (t,\vec{x}) \overset{T}{\longmapsto} (-t,\vec{x}).
\end{align}
But the Lagrangian of Eq.~\eqref{eq:fullQCDLag} is also invariant under each of these transformations separately.
If nuclear physics was only governed by the strong interactions, these symmetries should be reflected in the interactions between hadrons, such as the nuclear forces between protons and neutrons.
But these hadronic interactions are manifestations of the interactions between the quarks that make up the hadrons, and quarks also interact via the electromagnetic and weak interactions. 
Quantum electrodynamics (QED) as the field theory describing the electromagnetic interactions is also invariant under $C$, $P$, and $T$ separately, which means that the electromagnetic interactions between quarks do not break $C$, $P$, and $T$.
The situation is different for the weak interactions. 
The weak interactions violate $P$, $C$, and combined $CP$ (equivalent to $T$ because of $CPT$ invariance) symmetries. 
Therefore, there are components in the hadronic interactions that violate $P$ and $CP$ invariance.
However, since they originate from the weak interaction, they are expected to be much smaller than the symmetry-preserving components.
For example, there are parity-violating components of the nucleon-nucleon interactions that are expected to be suppressed by factors of about $10^{-7}$ relative to the parity-conserving interactions, see ~\citep{EncycPV} as well as, e.g.,~\citep{Ramsey-Musolf:2006vfz,Haxton:2013aca,Schindler:2013yua} for reviews.
CP-violating components are even more strongly suppressed, by many orders of magnitude.
So for much of nuclear physics, the discrete symmetries of $C$, $P$, and $T$ can be assumed to hold to a very good approximation.
Nonetheless, studying symmetry-violating effects can serve as an important tool for precision physics and in the search for new physics beyond the standard model of particle physics.
And because the symmetry-breaking effects are so small, they can be treated as perturbations to the dominant symmetry-conserving interactions.

The discussion above assumed that the strong interactions are governed by the Lagrangian of Eq.~\eqref{eq:fullQCDLag}, with $P$ and $CP$ nonconservation only originating in the weak interactions.
But $\SU{3}_c$ color gauge symmetry permits another term of mass dimension 4, the so-called $\theta$ term of QCD,
\begin{align}
\label{eq:theta}
    \calL_\theta  =\frac{g^2_s\bar{\theta}}{32\pi^2}
    \epsilon_{\mu\nu\rho\sigma} \text{Tr}_c
    \left( \calG^{\mu\nu} \calG^{\rho\sigma}\right),\quad \epsilon_{0123}=1.
\end{align}
This term explicitly breaks $P$ and combined $CP$ invariance. Current experimental information indicates that the $\theta$ term is highly suppressed, see, e.g.,~\citep{Kim:2008hd,Hook:2018dlk,Bonanno:2025wcv} for reviews.
It will therefore not be considered in the following discussion.

\begin{BoxTypeA}[box:discrete]{Discrete symmetries of QCD}

In the absence of the $\theta$ term, QCD is invariant under charge conjugation $C$, parity $P$, and time reversal $T$.

\end{BoxTypeA}

\section{Baryon number conservation}
\label{sec:BaryonNumber}

Hadrons, i.e., particles subject to the strong interactions, are divided into two categories: mesons, which in a simple quark-model picture consist of a quark and an antiquark, and baryons, which in the quark-model picture consist of three quarks. 
Despite hadrons being very complex composite states of quarks and gluons, these very simple (anti)quark assignments, also referred to as valence quarks, capture the main quantum numbers, e.g., electric charge and spin, of the hadrons.
For each particle there is a corresponding antiparticle.
Antimesons again consist of quark-antiquark pairs, while antibaryons consist of three antiquarks.
In any process observed so far, the total of the number of baryons minus the number of antibaryons remains the same.
This can be expressed as the conservation of a quantum number called baryon number $B$.
If we assign $B=1$ to each baryon, $B=-1$ to each antibaryon, and baryon number $B=0$ to each meson (and all other particles like electrons, etc), the total $B$ for the initial state and the total $B$ for the final state have to be the same. 
This does \emph{not} mean that the number of particles has to stay constant.
For example, the process $p + \pi^- \to n + \pi^0 + \pi^0$ is allowed, since the additional meson in the final state has $B=0$ and does not change the overall baryon number.
On the other hand, a process like proton decay, e.g., through $p\to e^+ + \pi^0$, is forbidden, since it changes baryon number from $B=1$ to $B=0$.

According to Noether's theorem \citep{Noether:1918zz}, a continuous symmetry is related to a conserved quantity.
For baryon number conservation we can indeed identify such a symmetry from the QCD Lagrangian.
Equation~\eqref{eq:QCDLagLight} remains invariant under multiplication of the quark fields by a common phase factor (a $\text{U}(1)$ transformation), 
\begin{align}
\label{eq:Btransform}
    q\mapsto e^{-i\theta}q.
\end{align}
The conserved quantity expected from Noether's theorem is
\begin{align}
    Q= \int d^3 x \ q^\dagger q.
\end{align}
This operator counts the total number of quarks minus the total number of antiquarks.
But this is just baryon number in a slight disguise.
If we assign baryon number $B=\frac{1}{3}$ to quarks and $B=-\frac{1}{3}$ to antiquarks, the conservation of $Q$ is equivalent to baryon number conservation.
We have thus identified the source of baryon number conservation as a symmetry of QCD.

The transformation of Eq.~\eqref{eq:Btransform} that leads to baryon number conservation multiplies all quark flavors by the same phase factor.
QCD is actually invariant under separate transformations of each individual quark flavor, i.e., the phase for $u$ quarks can be different from that for $d$ quarks, etc. 
As a result, in analogy to baryon number, we could assign up, down, and strange numbers, and each of these is conserved in QCD. 
This is less relevant than baryon number because quarks also interact through the weak interactions, and the weak interactions can change quark flavor. 
For example, a $d$ quark can convert into a $u$ quark with the emission of a $W$ boson, $d\to u + W^-$, thus changing both $d$ number and $u$ number. This process underlies the $\beta$ decay of a neutron into a proton.
While individual quark flavors are not conserved by the weak interactions, baryon number remains conserved.

Baryon number conservation is a property of the Lagrangian. 
In the standard model, baryon number conservation can be violated by nonperturbative effects of the weak interactions.
However, this nonconservation is extremely small at low temperatures.
Because baryon number is such a good symmetry of the standard model (e.g., the current limit on the proton lifetime is $\tau> \SI{9d29}{years}$ \citep{PDG}), the observation of a process in which baryon number is not conserved would be a signal of new physics beyond the standard model, see, e.g,~\citep{FileviezPerez:2022ypk,Broussard:2025opd} for recent reviews.
Baryon number nonconservation is also one of the Sakharov criteria for the observed asymmetry of matter over antimatter in the universe \citep{Sakharov:1967dj}.

\begin{BoxTypeA}[box:Bnumber]{Baryon number conservation in QCD}

If each baryon is assigned baryon number $B=1$, each antibaryon $B=-1$, and each meson $B=0$,  the total baryon number does not change in any QCD process.

\end{BoxTypeA}

\section{Isospin and flavor symmetry}
\label{sec:isospin}

Isospin was originally proposed as a symmetry involving protons and neutrons, based on the observation that these particles have nearly identical masses and can be considered as two states of a single particle, the nucleon \citep{Heisenberg:iso}. 
Isospin symmetry was then extended as an organizing principle for other hadrons, e.g., for the triplet of pions ($\pi^+,\pi^0,\pi^-$) or the quartet of $\Delta$ isobars ($\Delta^{++},\Delta^{+},\Delta^0,\Delta^-$).
It is now understood that isospin symmetry at the hadronic level originates in the isospin symmetry of the QCD Lagrangian. 

Consider the two-flavor case of Eq.~\eqref{eq:QCDLagLight}. Under transformations of the quarks $q$ with a \emph{matrix} $V \in \SU{2}$,
\begin{align}
\label{eq:isotrans}
	q \mapsto V q,
\end{align}
the Lagrangian transforms as
\begin{align}
    \calL_\text{QCD}^l &\mapsto \bar{q} V^\dagger \left( i\slashed{D}-\calM \right) V q -\frac{1}{2}\text{Tr}_c\left(\calG_{\mu\nu}\calG^{\mu\nu}\right).
\end{align}
The covariant derivative is independent of quark flavor, which means that it commutes with the matrix $V$. 
Since the unitary matrix $V$ satisfies $V^\dagger V = \id$, the term $\bar{q}i\slashed{D} q$ is invariant under the transformation of Eq.~\eqref{eq:isotrans}.
For the analysis of the mass term, it is convenient to write the $2\times 2$ version of the mass matrix of Eq.~\eqref{eq:qMassMat} as
\begin{align}
\label{eq:QCDMassDecompose}
    \calM= \begin{pmatrix} m_u & 0 \\
	0 & m_d \end{pmatrix} = \frac{1}{2}(m_u + m_d) \begin{pmatrix} 1 & 0 \\
	0 & 1 \end{pmatrix} + \frac{1}{2}(m_u - m_d) \begin{pmatrix} 1 & 0 \\
	0 & -1 \end{pmatrix} = \frac{1}{2}(m_u + m_d) \id + \frac{1}{2}(m_u - m_d) \tau_3,
\end{align}
where $\tau_3$ is the third Pauli matrix. The notation $\tau$ is chosen instead of the more common $\sigma$ to distinguish it as acting in the space of $u$ and $d$ quarks (i.e., isospin space) rather than spin space.
If $m_u=m_d$, the second term on the right-hand side vanishes and the mass matrix is proportional to the $2\times 2$ identity matrix.
Because the identity matrix commutes with any $2\times 2$ matrix, the full Lagrangian of Eq.~\eqref{eq:QCDLagLight} becomes invariant under   the transformations of Eq.~\eqref{eq:isotrans}.
This symmetry is referred to as the isospin symmetry of the strong interactions, and it requires that the $u$ and $d$ quark masses are equal.

Protons and neutrons can be treated analogously  to the $u$ and $d$ quarks. 
Collecting them in a two-component nucleon field $N$, 
\begin{align}
    N=
    \begin{pmatrix}
        p \\n
    \end{pmatrix},
\end{align}
this doublet of states transforms like the quark fields under an isospin transformation $V\in \SU{2}$,
\begin{align}
    N \mapsto V N.
\end{align}
In Lagrangians, the nucleon fields appear in so-called bilinears of the form $N^\dagger O_\tau N$, where $O_\tau$ denotes a $2 \times 2$ matrix (an operator) acting in isospin space.
These bilinears then transform as
\begin{align}
    N^\dagger O_\tau N \mapsto N^\dagger V^\dagger O_\tau V N
\end{align}
Unless $O_\tau$ is proportional to the identity matrix in isospin space $\id_\tau$, it will not commute with all possible isospin transformations $V$. 
This means that $N^\dagger \id_\tau N$, is the only bilinear that is invariant under isospin transformations by itself.
However, one can form combinations of two or more bilinears that are invariant, even though individual bilinears are not.
For example, a term like 
\begin{align}
\label{eq:NNiso}
    \sum_{a=1}^3(N^\dagger \tau_a N)(N^\dagger \tau_a N)
\end{align}
is invariant under isospin transformations, while a single factor $N^\dagger \tau_a N$ is not. 
Terms like the one in Eq.~\eqref{eq:NNiso} contribute to the \NN interactions.

Isospin invariance requires $m_u = m_d$.
But as seen in Table~\ref{tab:qmasses}, the $u$ and $d$ masses are not equal.
If $m_u\ne m_d$, the second term on the right-hand side of Eq.~\eqref{eq:QCDMassDecompose} cannot be neglected, the mass matrix no longer commutes with a general \SU{2} matrix, and isospin symmetry is broken. 
Naively, one would expect that the size of isospin breaking effects is related to the relative sizes of the isospin-symmetric and isospin-breaking terms in Eq.~\eqref{eq:QCDMassDecompose}, i.e., to the ratio $(m_u-m_d)/(m_u+m_d)$.
This ratio is not particularly small for the quark masses given in Table \ref{tab:qmasses}, suggesting that isospin should not be a good symmetry, contrary to observations. 
However, due to the nonperturbative nature of QCD, the ratio governing the impact of different quark masses is $(m_u-m_d)/\Lamqcd$, with \Lamqcd an intrinsic scale of QCD on the order of $200-500\,\text{MeV}$, meaning that even at the lower end of this range isospin is broken at the order of 1\%. 
In addition to the difference in the quark masses, isospin symmetry is also broken by electromagnetic effects, since the $u$ and $d$ quarks have different electric charges, but again these effects are typically very small, on the order of a few percent. 
So while isospin is not an exact symmetry of QCD, it is still an approximate one, and it is still very useful in nuclear physics.
It is typically a good approximation to start in the isospin-symmetric limit and include isospin-breaking effects as perturbations.

These considerations can be extended to three flavors.
As seen from Table~\ref{tab:qmasses}, the $s$ quark mass is significantly larger than the $u$ and $d$ masses. 
Nonetheless, the ratio $(m_s-m_{u/d})/\Lamqcd < 1$.
If $m_u=m_d=m_s$, the mass matrix in Eq.~\eqref{eq:qMassMat} becomes proportional to the $3\times 3$ identity matrix, and the three-flavor Lagrangian is again invariant under transformations of the quark fields, this time with an \SU{3} matrix.
The resulting symmetry is an \SU{3} flavor symmetry, which was originally proposed by Gell-Mann and Ne'eman \citep{Gell-Mann:1961omu,Neeman:1961jhl}. 
It is again an approximate symmetry, but the size of the $s$ quark mass means that SU(3) flavor symmetry breaking is much more significant than isospin breaking.

\begin{BoxTypeA}[box:isospin]{Isospin symmetry of QCD}

Isospin symmetry is an approximate symmetry of the QCD Lagrangian under unitary transformations in the space of $u$ and $d$ quarks. It is an exact symmetry in the limit that the $u$ and $d$ quark masses are equal. It is broken by the quark mass difference and by electromagnetic effects. However, these isospin-breaking effects are very small and typically can be treated perturbatively.

\end{BoxTypeA}

\section{Symmetries of massless QCD}
\label{sec:SymMassless}

As seen in the previous section, assuming that the $u$ and $d$ masses are equal leads to isospin as an additional symmetry of QCD. 
And while it is only an approximate symmetry, isospin is still useful as the leading order (LO) in a perturbative treatment.
Not only are the $u$ and $d$ quark masses very similar, the light quark masses are also significantly smaller than any hadronic masses, with the exception of pions. 
For example, in a simple quark model picture, a proton consists of two $u$ and one $d$ quark. 
The sum of these quark masses of approximately \SI{9}{MeV} is tiny when compared to the proton mass of approximately \SI{938}{MeV}.
Thus, it does not seem unreasonable to set the light quark masses to zero to see if any further insight can be gained from the QCD Lagrangian. 
Taking the light quark masses $m_l\to 0$ is referred to as the chiral limit. 
One distinguishes between the three-flavor (or $\SU{3}$ - this nomenclature will become clearer below) chiral limit, in which all three masses $m_u, m_d, m_s$ are assumed to be zero, and the two-flavor (or $\SU{2}$) chiral limit, in which $m_u$ and $m_d$ are zero, but $m_s$ is kept at its physical value.
In the chiral limit, denoted by a superscript 0, the Lagrangian of Eq.~\eqref{eq:QCDLagLight} simplifies to
\begin{align}
\label{eq:chiralQCDLag}
    \calL_\text{QCD}^0 = \bar{q} i\slashed{D} q -\frac{1}{2}\text{Tr}_c\left(\calG_{\mu\nu}\calG^{\mu\nu}\right).
\end{align}
Compared to the Lagrangian with finite quark masses, the Lagrangian of Eq.~\eqref{eq:chiralQCDLag} possesses additional symmetries. 
To make these more apparent, it is convenient to introduce left- and right-handed quark fields,
\begin{align}
\label{eq:LRfields}
	q_L = \frac{1}{2} (\mathbbm{1}-\gamma_5) q ,\quad q_R =  \frac{1}{2} (\mathbbm{1}+\gamma_5) q,
\end{align}
where $\gamma_5 = i\gamma^0\gamma^1\gamma^2\gamma^3$, with the $\gamma^i$ the Dirac matrices. 
The massless QCD Lagrangian then takes the form
\begin{align}
\label{eq:chiralQCDLagLR}
    \calL_\text{QCD}^0 = \bar{q}_{L} i\slashed{D} q_L + \bar{q}_{R} i\slashed{D} q_R -\frac{1}{2}\text{Tr}_c\left(\calG_{\mu\nu}\calG^{\mu\nu}\right).
\end{align}
This form shows that in the chiral (massless) limit, the left-handed quark fields only couple to left-handed quark fields, and right-handed fields only to right-handed fields.  
We now consider \emph{independent} unitary transformations of the left- and right-handed fields,
\begin{align}
\label{eq:ChiralQuarkTrans}
	q_L \mapsto \tilde{U}_L q_L, \quad q_r \mapsto \tilde{U}_R q_R, \quad \tilde{U}_{L/R} \in \text{U}(N)_{L/R}, 
\end{align}
under which
\begin{align}
    \calL_\text{QCD}^0 \mapsto \calL_\text{QCD}^{0^{\, \prime}} & =\bar{q}_{L} \tilde{U}_L^\dagger i\slashed{D} \tilde{U}_L q_L + \bar{q}_{R} \tilde{U}_R^\dagger i\slashed{D} \tilde{U}_R q_R -\frac{1}{2}\text{Tr}_c\left(\calG_{\mu\nu}\calG^{\mu\nu}\right) \notag \\
    & = \bar{q}_{L} i\slashed{D} q_L + \bar{q}_{R} i\slashed{D} q_R -\frac{1}{2}\text{Tr}_c\left(\calG_{\mu\nu}\calG^{\mu\nu}\right) \\
    & = \calL_\text{QCD}^0 \notag,
\end{align}
i.e., the Lagrangian is invariant under these transformations. 
Here we have used that the covariant derivative is independent of the quark flavors and commutes with the unitary matrices $\tilde{U}_{L/R}$.
We thus see that the QCD Lagrangian in the chiral limit has a global $\text{U}(N)_L\times \text{U}(N)_R$ symmetry, where $N=2,3$ depending on which chiral limit is considered. 
It is convenient to decompose the $\text{U}(N)_L\times \text{U}(N)_R$ transformations into $\SU{N}_L \times \SU{N}_R \times \text{U}(1)_L \times \text{U}(1)_R$ transformations:
\begin{align}
    \tilde{U}_{L/R} = e^{-i\theta_{L/R}} U_{L/R}, \quad U_{L/R} \in \SU{N}_{L/R}.
\end{align}
The $\text{U}(1)_L \times \text{U}(1)_R$ transformations correspond to multiplication of the left- and right-handed quark fields by independent phase factors,
\begin{align}
    q_L \mapsto e^{-i\theta_L} q_L, \quad q_R \mapsto e^{-i\theta_R} q_R.
\end{align}
Equivalently, one can consider $\text{U}(1)_V \times \text{U}(1)_A$ transformations (the $V$ stands for vector and the $A$ for axial-vector),
\begin{alignat}{2}
    \text{U}(1)_V: \quad q_L & \mapsto e^{-i\theta_V} q_L, &\quad  q_R &\mapsto e^{-i\theta_V} q_R, \\
    \text{U}(1)_A: \quad q_L &\mapsto e^{i\theta_A} q_L, & q_R &\mapsto e^{-i\theta_A} q_R.
\end{alignat}
The $\text{U}(1)_V$ symmetry corresponds to the symmetry discussed in Sec.~\ref{sec:BaryonNumber}, in which left- and right-handed fields are multiplied by the same phase factor (with $\theta_V = \theta$). 
It is responsible for baryon number conservation and holds also for finite quark masses.
The new feature in the chiral limit is the invariance of the massless QCD Lagrangian under $\SU{N}_L \times \SU{N}_R \times \text{U}(1)_A$ transformations.

\subsection{Axial anomaly}
\label{sec:anomaly}

The invariance under axial $\text{U}(1)_A$ transformations is a symmetry of the massless QCD Lagrangian.
It is broken explicitly by finite quark masses.
In the case of isospin symmetry in Sec.~\ref{sec:isospin} we showed that the mass difference between the $u$ and $d$ quarks breaks isospin symmetry explicitly, but argued that these effects are small and can be treated in perturbation theory.
One could thus ask whether an analogous argument holds for the axial $\text{U}(1)_A$:
Since $m_u, m_d \ll \Lamqcd$, is the invariance under $\text{U}(1)_A$ an approximate symmetry?

The answer is no, and even for massless quarks the $\text{U}(1)_A$ symmetry is broken. 
So far we have only considered the \emph{form} of the Lagrangian and our arguments have been purely classical. 
In the case of $\text{U}(1)_A$, quantum effects break the symmetry \citep{Adler:1969gk,Adler:1969er,Bell:1969ts}.
This is referred to as anomalous symmetry breaking.
While the axial anomaly is very important and interesting, from now on we will  simply use that the $\text{U}(1)_A$ symmetry is broken and not a symmetry of the QCD Lagrangian.

\subsection{Chiral symmetry}
The remaining $\SU{N}_L\times \SU{N}_R$ symmetry is referred to as the chiral symmetry of massless QCD. 
It corresponds to the invariance of the QCD Lagrangian with massless quarks under independent \SU{N} ($N=2,3$) transformations of left- and right-handed quark fields:
\begin{align}
    q_L\mapsto U_L q_L,\quad q_R \mapsto U_R q_R.
\end{align}
Chiral symmetry is a global symmetry, i.e., while the quark fields are functions of $x$, the matrices $U_{L/R}$ are not. 
Chiral symmetry plays a very important role in describing the strong interactions for processes at energies well below \SI{1}{GeV}. 
However, as seen above for the anomaly, simply considering the symmetries of the Lagrangian might not be sufficient. We also still need to address the fact that the quark masses are not actually zero.

\begin{BoxTypeA}[box:SymQCD]{Chiral symmetry of QCD}

Invariance of the QCD Lagrangian with $N$ massless flavors under independent $\SU{N}_L\times \SU{N}_R$ transformations of left- and right-handed quark fields.

\end{BoxTypeA}

\section{Chiral symmetry breaking}

The discussion of chiral symmetry so far was based on the form of the Lagrangian  in the limit of the light quark masses being zero. 
However, our current understanding is that the quark masses are \emph{not} zero, see Table \ref{tab:qmasses}.
As discussed in Sec.~\ref{sec:Explicit} below, these nonzero masses break chiral symmetry. 
But even for zero quark masses, the situation is not as simple as the discussion of the form of the Lagrangian seems to suggest. 
The symmetries of the Lagrangian should be reflected in what can be observed in experiments, e.g., through conservation laws or relationships between particle masses.
As we will now see, the situation for QCD and chiral symmetry is more interesting than naively expected. 
There are strong indications that, even for zero quark masses, chiral symmetry is broken spontaneously.

Chiral symmetry is thus broken in two different ways: spontaneously even for zero quark masses, and by the presence of nonzero quark masses.
Nonetheless, the concept of chiral symmetry and, in fact, the particular forms of its breaking, have been very influential for low-energy nuclear physics.

\subsection{Spontaneous chiral symmetry breaking}
\label{sec:SSB}

One might expect that the symmetries of a Lagrangian should also be reflected in the symmetries of the corresponding spectrum of particles. 
In particular, for a continous symmetry such as an internal \SU{N} invariance, there should be groups of particles with equal mass, referred to as multiplets. 
Mathematically, these correspond to irreducible representations of the symmetry group in question.
However, according to a theorem by \citet{10.1063/1.1931207}, the symmetry of the spectrum is related to the symmetry of the ground state of a theory.
It is possible that the symmetry of the ground state is \emph{not} the same as that of the Lagrangian. 
If the ground state is not invariant under the full symmetry group of the Lagrangian, the symmetry is said to be spontaneously broken. 
There are many examples of spontaneous symmetry breaking in physics, e.g., it plays an important role in the Higgs mechanism \citep{Englert:1964et,Higgs:1964pj,Guralnik:1964eu}.

How does this relate to QCD and chiral symmetry? The symmetries of the QCD Lagrangian should be reflected in the spectrum of strongly-interacting particles, the hadrons.
In particular, invariance under $\SU{N}_L\times \SU{N}_R$ ($N=2,3$) would imply that for each $\SU{N}$ multiplet there is a second $\SU{N}$ multiplet of equal mass, but opposite parity. 
However, this so-called parity doubling is not observed among light hadrons. While hadrons can (approximately) be arranged in terms of \SU{N} multiplets, the multiplets of opposite parity have very different masses.
This suggests that only a single SU($N$) symmetry is reflected in the hadron spectrum. 
Thus, the ground state of QCD does not appear to have the same symmetry as the QCD Lagrangian, and it is assumed that chiral symmetry is spontaneously broken from $\SU{N}_L\times\SU{N}_R$ to \SU{N}.

\begin{figure}[t]
\centering
    \includegraphics[width=.5\textwidth]{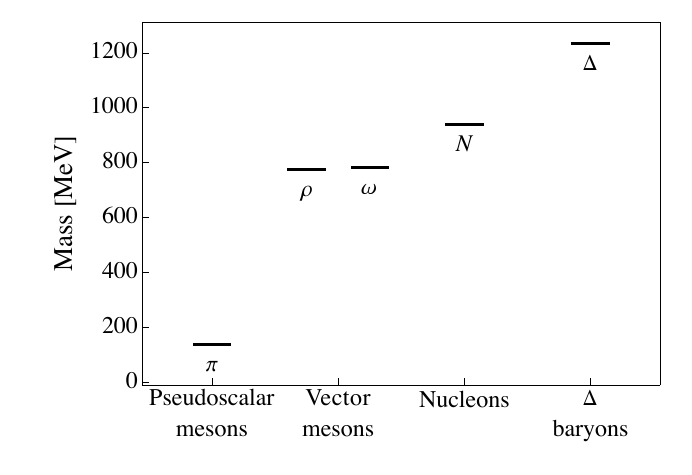}
\caption{Masses of light hadrons with only $u$ and $d$ valence quarks.} \label{fig:hadmass}
\end{figure}

There is another indication that chiral symmetry is spontaneously broken. 
The spontaneous breaking of a continuous symmetry is accompanied by the emergence of massless particles, the so-called Goldstone bosons \citep{Nambu:1960tm,Goldstone:1961eq}. 
The number of expected Goldstone bosons is related to the underlying symmetry group of the Lagrangian and the symmetry of the ground state/spectrum. For $\SU{2}_L\times \SU{2}_R$ being spontaneously broken to \SU{2}, one expects three Goldstone bosons, while for the $\SU{3}_L\times \SU{3}_R \to \SU{3}$ case the number of Goldstone bosons is eight. 
The triplet of pions (with only $u$ and $d$ valence quarks, i.e., the $N=2$ case) is much lighter than other hadrons that do not contain any $s$ valence quarks, see Fig.~\ref{fig:hadmass}.
Similarly, the lowest-lying octet of pseudoscalar mesons (with valence $s$ quarks, i.e., $N=3$) is significantly lighter than other hadrons.\footnote{For a discussion of spontaneous chiral symmetry breaking from a lattice QCD perspective, see, e.g., \cite{Faber:2017alm} and references therein.} 
Even though they are not massless, these light mesons are identified with the Goldstone bosons expected from spontaneous symmetry breaking. Their nonzero masses originate from the nonzero quark masses that we have neglected up to this point.

\subsection{Explicit chiral symmetry breaking}
\label{sec:Explicit}

The discussion of chiral symmetry up to this point has focused on \emph{massless} quarks. 
However, in the real world the quark masses are not identically zero. In terms of the left- and right-handed fields of Eq.~\eqref{eq:LRfields} the mass term is given by 
\begin{align}
\label{eq:QCDMassTerm}
	\calL_{\calM} = -\bar{q} \calM q =  -\bar{q}_R \calM q_L - \bar{q}_L \calM^\dagger q_R,
\end{align}
with the quark mass matrix defined in Eq.~\eqref{eq:qMassMat}.
The quark mass matrix is real; the Hermitian conjugate in the last-term of Eq.~\eqref{eq:QCDMassTerm} is included for later convenience.
Unlike the massless Lagrangian of Eq.~\eqref{eq:chiralQCDLagLR}, the mass term couples left-handed and right-handed fields. 
Performing a chiral transformation, it takes the form
\begin{align}
\label{eq:QCDMassTermTrans}
    \calL_{\calM} \mapsto -\bar{q}_R U_R^\dagger \calM U_L q_L - \bar{q}_L U_L^\dagger\calM^\dagger U_R q_R
\end{align}
Even if the mass matrix commuted with $U_{L/R}$ (and in general it does not), this term is \emph{not} invariant under independent transformations of the left- and right-handed fields because in general $U_L^\dagger U_R \ne \id$. In other words, the mass term breaks chiral symmetry. This breaking is referred to as an explicit symmetry breaking, as it can be read off directly  from the form of the mass term in the Lagrangian.

\begin{BoxTypeA}[box:ChiralsymBreaking]{Chiral symmetry breaking in QCD}

It is assumed that the chiral $\SU{N}_L\times \SU{N}_R$ symmetry is spontaneously broken to $\SU{N}$, leading to the appearance of Goldstone bosons, which are identified with the lightest pseudoscalar mesons (the pions in the two-flavor case). In addition, the nonzero physical quark masses break chiral symmetry explicitly, leading to nonzero masses for the Goldstone bosons.

\end{BoxTypeA}

\section{Effective field theory}
\label{sec:EFT}

If chiral symmetry is broken both spontaneously and explicitly and is thus not actually a symmetry of QCD, what is its relevance for nuclear physics?
While chiral symmetry is not an exact symmetry of QCD, the quark masses responsible for explicit chiral symmetry breaking are much smaller than the intrinsic QCD scale $\Lambda_\text{QCD}$ and other hadronic mass scales, such as, e.g., the masses of the $\rho$ mesons, $M_\rho \approx \SI{775}{MeV}$. 
This leads to the idea that chiral symmetry is an approximate symmetry of QCD, similar to isospin symmetry discussed above. Further, it might be possible that the symmetry breaking effects due to the nonzero quark masses can be treated as perturbative corrections.

Further, not only the exact symmetries of QCD, but also the patterns of any symmetry breaking should be reflected in low-energy hadronic physics.
The presence of Goldstone bosons is a result of spontaneous chiral symmetry breaking, but spontaneous chiral symmetry breaking implies even more than that: it also imposes constraints on how the Goldstone bosons can interact with each other.
The modern formulation of these ideas is in the form of an effective field theory (EFT) and goes back to \citet{Weinberg:1978kz}. 
The underlying principle is often referred to as Weinberg's folk theorem:
\begin{quote}
[I]f one writes down the most general possible Lagrangian, including all terms consistent with the assumed symmetry principles, and then calculates matrix elements with this Lagrangian to any given order of perturbation theory, the result will simply be the most general possible S-matrix consistent with analyticity, perturbative unitarity, cluster decomposition, and the assumed symmetry properties.
\end{quote}

What does this mean for QCD? The first step is to choose which degrees of freedom to use in our calculations. 
While QCD is formulated in terms of quarks and gluons, at energies well below \SI{1}{GeV} (referred to as low energies from now on) it is more convenient to work with the hadrons observed in experiments, like pions and nucleons.
The next step is to understand how these degrees of freedom transform under the (approximate) symmetries of the QCD Lagrangian, in particular under the $\SU{N}_L\times\SU{N}_R$ chiral symmetry.
It is then possible to construct the most general Lagrangian in terms of these degrees of freedom that is ``consistent with the assumed symmetry principles.''
This Lagrangian is referred to as an effective Lagrangian.
In general, however, this effective Lagrangian will consist of an \emph{infinite} number of terms, and each term in the Lagrangian comes with a coefficient, a so-called low-energy constant (LEC), that somehow needs to be determined before testable predictions can be made.
This is not a useful approach unless there is some way to organize all these terms and how they contribute to observables.

The solution to this problem lies in the application of perturbation theory. 
Performing a perturbative calculation requires the existence of some small expansion parameter(s). 
When perturbation theory is first introduced in quantum mechanics lectures, this parameter is typically some (dimensionless) coupling constant.
This approach is not suitable for QCD at low energies: its coupling is not small. 
However, quantities other than coupling constants can serve as small expansion parameters.
The multipole expansion in electrodynamics serves as an example. 
When considering a static charge distribution that is localized in a region of radius $r'$, the resulting potential at a point a distance $r$ outside this region can be expanded in a series of potentials due to electrostatic moments, e.g., the monopole moment (the total charge), the dipole moment, etc. 
The expansion parameter in this case is given by the ratio $\frac{r'}{r}$.
Formally, the expansion contains an infinite number of terms. 
But it is useful as an approximation: truncating the expansion after a finite number of terms provides an approximation of the exact potential of the charge distribution as long as $\frac{r'}{r} \ll 1$. 
By including higher-order terms, the approximation can be improved.

Since the QCD coupling cannot be used a small expansion parameter, we must look for some other small parameters instead, and the quark masses seem like a good starting point.
However, the idea of ``small'' is meaningless for a dimensionful quantity by itself: it is only small relative to some particular scale. 
This is also the case in the multipole expansion: it is the \emph{ratio} of the two length scales $r'$ and $r$ that forms the small expansion parameter, not one individual length.
In the case of the multipole expansion, this ratio emerges from expanding the known exact expression for the potential.
For low-energy QCD, however, we do not have any analytic solutions which we could expand. 
So if the quark masses enter the ratio that forms our expansion parameter in the numerator, what is the denominator?
As mentioned above, QCD has an intrinsic scale \Lamqcd that could serve as the denominator, and $m_{u,d}/\Lamqcd \ll 1$.
There could also be additional factors of order 1, so for now we will simply assume that a scale $\Lambda\gg m_{u,d}$ exists.
The method to determine which terms from the infinite number of terms in the effective Lagrangian contribute to an observable at a given order in the small-parameter expansion is called \emph{power counting}, which will be discussed in more detail below.

\begin{BoxTypeA}[box:EFT]{Effective field theory}

An EFT is a low-energy approximation to an underlying theory, based on symmetries and a separation of energy scales. It is formulated in terms of the degrees of freedom relevant at the considered energies. An associated power counting provides a method to assess the relative importance of the various terms appearing in the EFT.

\end{BoxTypeA}

\section{Meson chiral perturbation theory}
\label{sec:mesonchpt}

We now apply the concepts of the previous section to QCD at low energies, starting with a discussion of the interactions of the Goldstone bosons arising from the spontaneous breaking of chiral symmetry. 
This includes both the construction of the effective Lagrangian and the development of a power counting.
The resulting theory is referred to as chiral perturbation theory (\chipt).

\subsection{Leading-order Lagrangian for pions in the chiral limit}

We will start with the two-flavor case of assuming that the $u$ and $d$ quark masses are zero (and keeping the $s$ quark mass at its physical value). 
The QCD Lagrangian has an $\SU{2}_L\times \SU{2}_R$ chiral symmetry that is spontaneously broken to the isospin group $\SU{2}_V$.
As a result, there are three Goldstone bosons, which we identify with the pions $\pi^+, \pi^0,\pi^-$.
We will use these as the degrees of freedom in our effective Lagrangian.
Spontaneous symmetry breaking also dictates how the Goldstone bosons transform under the symmetry \citep{Weinberg:1968de,Coleman:1969sm,Callan:1969sn}.
There are different ways to parametrize the Goldstone boson fields and how they transform. 
Here, we will use an exponential parametrization, collecting the Goldstone bosons in an \SU{2} matrix $U(x)$,
\begin{align}
\label{eq:Udef2}
    U(x) = \exp\left[i\frac{\phi(x)}{F} \right],
\end{align}
where $\phi(x)$ is a $2\times 2$ matrix containing the pion fields (suppressing the $x$ dependence),
\begin{align}
\label{eq:pimatrix}
    \phi & 
    =
    \begin{pmatrix}
        \pi^0 & \sqrt{2} \pi^+ \\
        \sqrt{2} \pi^- & -\pi^0
    \end{pmatrix}= \sum_{a=1}^3 \phi_a \tau_a.
\end{align}
The last expression is often more convenient for actual calculations.
The $\tau_a$ are the Pauli matrices and the fields $\phi_a$ are the Cartesian components of the pion fields; they are related to the physical pion fields by 
\begin{align}
\label{eq:PiPhysCart}
    \pi^{\pm} = \frac{1}{\sqrt{2}}(\phi_1 \mp i\phi_2), \quad \pi^0= \phi_3.
\end{align}
The pion fields have dimensions of energy, which can be seen as follows: 
A Lagrangian density has dimensions of four powers of $[\text{energy}]$ (recall that we are using units in which $c=1$, so that energy, momentum, and mass have the same dimensions). 
The kinetic term for pions is of the form $\frac{1}{2}\partial_\mu \phi_a \partial^\mu \phi_a$, and each derivative corresponds to a factor of dimension $[\text{energy}]$. 
Thus each factor of the pion field must also have dimensions of $[\text{energy}]$.
The constant $F$ in Eq.~\eqref{eq:Udef2} is introduced to make the exponent dimensionless, it thus also has dimensions of energy.
As we illustrate in Sec.~\ref{sec:local}, it is the pion decay constant in the chiral limit.
The physical value of the pion decay constant, i.e., its value at the physical quark masses, is given by $F_\pi \simeq \SI{92}{MeV}$.
The difference between the chiral-limit and the physical values can be calculated in \chipt \citep{Gasser:1983yg}.\footnote{In other conventions, the definition of $F$ can differ from the one given here by a factor of $\sqrt{2}$, resulting in a value of $\approx\SI{132}{MeV}$.}
Being an exponential matrix, the Goldstone boson matrix $U$ does not contain just a single pion field, but an infinite number of terms with increasing numbers of pion fields,
\begin{align}
\label{eq:pimatrixExp}
    U = \id + i \frac{\phi}{F} -  \frac{\phi^2}{2F^2} +\cdots.
\end{align}
This means that relationships exist in the effective Lagrangian between terms with different numbers of pion fields.

Having identified the degrees of freedom and how to parametrize them, we need to understand their behavior under symmetry transformations. 
As mesons, the pions have baryon number 0 and transform trivially under the $U(1)_V$ symmetry of QCD, i.e., $\phi\mapsto \phi$, which implies $U\mapsto U$. 
Thus, this symmetry does not restrict the form of the effective Lagrangian.
Under chiral transformations $(L,R)\in \SU{2}_L\times \SU{2}_R$, the Goldstone boson matrix in our convention transforms as
\begin{align}
\label{eq:Utransform}
    U(x) \mapsto U'(x) = RU(x) L^\dagger.
\end{align}

We can now start to construct the most general effective Lagrangian. 
It cannot be a matrix, so the Lagrangian will involve traces over the $U$ matrices.
A term $\Tr(U)$ with a single factor of $U$ is not invariant under chiral transformations.
Similarly, a term with two factors of $U$ is not invariant.
The simplest chirally invariant term is
\begin{align}
    \Tr(U U^\dagger) = \Tr (U^\dagger U),
\end{align}
with the equality of the two terms a general property of the trace, $\Tr(AB) = \Tr (BA)$. 
The Hermitian conjugate is required for invariance under chiral transformations:
\begin{align}
    \Tr(U U^\dagger) \mapsto \Tr(RUL^\dagger LU^\dagger R^\dagger) =  \Tr(RUU^\dagger R^\dagger) = \Tr(R^\dagger R U U^\dagger) = \Tr(U U^\dagger),
\end{align}
where we have used that $L$ and $R$ are unitary matrices and again $\Tr(AB) = \Tr (BA)$.
Since $U$ is an \SU{2} matrix, $U U^\dagger =\id$ and 
\begin{align}
    \Tr(U U^\dagger) = 2.
\end{align}
A constant term in the Lagrangian does not affect the Euler-Lagrange equations and can therefore be ignored for our purposes. 
Analogous arguments hold for attempting to construct other terms that contain only $U$ and $U^\dagger$.
Chiral symmetry requires that they appear in combinations of $UU^\dagger$ or $U^\dagger U$, but the unitarity of the matrix $U$ means that these are just constant terms.
This illustrates a general result: interactions between the Goldstone bosons of spontaneous chiral symmetry breaking must contain derivatives of the fields.

\begin{figure}[t]
\centering
\begin{subfigure}[b]{0.3\textwidth}
\centering
     \includegraphics[width=.5\textwidth]{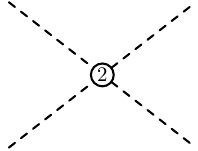}
    \caption{}\label{fig:PiPiLO}
\end{subfigure}
\begin{subfigure}[b]{0.3\textwidth}
\centering
    \includegraphics[width=.5\textwidth]{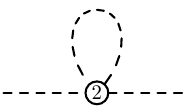}
    \caption{}\label{fig:PiMass}
\end{subfigure}
\caption{(a) Feynman diagram representing a four-pion vertex. (b) Loop diagram obtained from the vertex in (a) contributing to the pion mass.}
\end{figure}

Since $(L,R)$ represent \emph{global} transformations that are independent of $x$, derivatives of $U$ and $U^\dagger$ transform like the fields themselves,
\begin{align}
    \partial_\mu U \mapsto \partial_\mu (RUL^\dagger) = R(\partial_\mu U) L^\dagger, \quad \partial_\mu U^\dagger  \mapsto L(\partial_\mu U^\dagger) R^\dagger.
\end{align}
But because the effective Lagrangian is a Lorentz scalar, the Lorentz indices on the derivatives must be contracted with other Lorentz indices;
terms like 
\begin{align}
\label{eq:trdUU}
    \Tr(\partial_\mu U U^\dagger)
\end{align}
cannot appear by themselves.
Since (for now) derivatives are the only Lorentz vectors at our disposal, the effective Lagrangian will contain terms with an even number of derivatives.
The term of Eq.~\eqref{eq:trdUU} is actually identically zero, and the only linearly independent term with only two derivatives is given by
\begin{align}
\label{eq:twoder}
    \frac{F^2}{4}\Tr(\partial_\mu U \partial^\mu U^\dagger).
\end{align}
The factor of $F^2/4$ is the first example of a low-energy constant (LEC). 
Each independent term in the Lagrangian is accompanied by its own LEC. 
The LECs incorporate all the details of QCD that are not resolved in our effective description in terms of hadrons.
Since $U$ and $U^\dagger$ are dimensionless and each derivative corresponds to a term of dimensions of $[\text{energy}]$, the LEC for the term of Eq.~\eqref{eq:twoder} must have dimensions of two powers of $[\text{energy}]$.
The parametrization of the LEC is chosen to generate the standard kinetic term of a scalar field when $U$ is expanded in the number of pion fields as in Eq.~\eqref{eq:pimatrixExp},
\begin{align}
    \frac{F^2}{4}\Tr(\partial_\mu U \partial^\mu U^\dagger) = \frac{1}{2}\sum_{a=1}^3 \partial_\mu \phi_a \partial^\mu \phi_a + \cdots ,
\end{align}
where the ellipsis denotes terms with more pion fields.
For example, the next term in this expansion contains four pion fields and has the form
\begin{align}
\label{eq:twodertwopi}
    \frac{1}{6F^2} \sum_{a,b=1}^3 \left( \phi_a \partial_\mu \phi_a \partial^\mu \phi_b \phi_b - \phi_a \phi_a \partial_\mu \phi_b\partial^\mu \phi_b \right).
\end{align}
It corresponds to interactions represented by the vertex shown in Fig.~\ref{fig:PiPiLO}, where here the label 2 denotes the number of derivatives.
Since this diagram can be interpreted as having two incoming pion lines and two outgoing pion lines, this vertex contributes to $\pi\pi$ scattering (see below). 
But the same vertex also contributes to other processes.
For example, two of the pion lines can be connected with a pion propagator.
The result is a diagram with two external pion lines that contributes to the calculation of the pion mass (see, e.g.,~\citep{Scherer:2012xha} for details of how to perform this calculation). 

The requirement that Goldstone boson interactions must contain derivatives is crucial to organizing the infinite number of terms in the effective Lagrangian.
A derivative in the Lagrangian translates into a factor of momentum in the Feynman rules. 
As long as one restricts the discussion to energies and momenta that are much smaller than the scale $\Lambda$, each additional derivative will correspond to a suppression by a factor of $p/\Lambda$, where $p$ denotes a small momentum.
Terms with four derivatives, e.g.,
\begin{align}
    \Tr(\partial_\mu U \partial^\mu U^\dagger\partial_\nu U \partial^\nu U^\dagger),
\end{align}
are suppressed relative to the term in Eq.~\eqref{eq:twoder}.
Thus we have identified a second expansion parameter $p/\Lambda$ in addition to the quark mass terms $m_{u,d}/\Lambda$.

\subsection{Including finite quark masses}
\label{sec:ChPTmass}

How can we incorporate the explicit symmetry breaking due to the nonzero quark masses in the effective Lagrangian?
The QCD quark mass term is given in Eq.~\eqref{eq:QCDMassTerm},
\begin{align*}
	\calL_{\calM} =  -\bar{q}_R \calM q_L - \bar{q}_L \calM^\dagger q_R. \tag{\ref{eq:QCDMassTerm}}
\end{align*}
Our goal is to not only write down terms that break chiral symmetry, but to also capture \emph{how} the quark masses break the symmetry.
To do so, we observe that $\calL_{\calM}$ \emph{would be} invariant under chiral transformations \emph{if} $\calM$ transformed as
\begin{align}
\label{eq:Mtransform}
    \calM \mapsto R\calM L^\dagger, \quad \calM^\dagger \mapsto L\calM^\dagger R^\dagger.
\end{align}
This also explains why we use $\calM^\dagger$ in the second term of Eq.~\eqref{eq:QCDMassTerm} despite $\calM$ being a real, diagonal matrix; it ensures that the second term is also invariant under chiral transformations.
We can then use $\calM$ and its assumed transformation in combination with $U$ and its derivatives to construct terms contributing to the effective Lagrangian that are invariant under chiral transformations.
There are two such terms with a single factor of $\calM$,
\begin{align}
\label{eq:pmMass}
    \Tr(\calM U^\dagger \pm U \calM^\dagger).
\end{align}
This approach might seem a bit odd.
The matrix $\calM$ is a constant matrix and does not actually transform at all.
It does break chiral symmetry, but without any additional guidance we would not know how to incorporate that symmetry breaking into the effective Lagrangian.
Assuming the transformation behavior of Eq.~\eqref{eq:Mtransform} lets us incorporate not just the fact that the quark masses break chiral symmetry, but also the pattern of how they break it.
This is a \emph{purely theoretical} tool.
Once the Lagrangian is constructed, the constant quark mass matrix $\calM$ is used in calculations.
As a result, the terms in Eq.~\eqref{eq:pmMass} are \emph{not} invariant under chiral transformations, but they do encode the symmetry breaking of the QCD mass term of Eq.~\eqref{eq:QCDMassTerm}.

But chiral symmetry is not the only symmetry of QCD that we need to incorporate.
As discussed in Sec.~\ref{sec:discrete}, QCD is also invariant under parity.
At the level of the effective Lagrangian, the term in Eq.~\eqref{eq:pmMass} with the relative minus sign is odd under parity and can therefore be excluded.
The Lagrangian that incorporates explicit symmetry breaking and the discrete symmetries of QCD at lowest order in $\calM$ is therefore given by
\begin{align}
\label{eq:Lsb}
    \calL_{\text{s.b.}} = \frac{F^2 B}{2} \Tr(\calM U^\dagger + U \calM^\dagger).
\end{align}
The terms in the trace have dimensions of $[\text{energy}]$, so it must be multiplied by an LEC of dimensions $[\text{energy}]^3$.
The factor $F^2/2$ is included for future convenience, and since it has dimensions of $[\text{energy}]^2$, the additional LEC $B$ has dimensions of $[\text{energy}]$.

To understand the meaning of this term, we insert the expansion of the matrix $U$ of Eq.~\eqref{eq:pimatrixExp} into Eq.~\eqref{eq:Lsb}, which generates terms with $0,2,4,\ldots$ pion fields.\footnote{That there are only an even number of fields can be seen from a symmetry argument. Under $\phi\to -\phi$, the exponential $U\to U^\dagger$, and $\calL_{\text{s.b.}}$ is invariant under this transformation. The same argument applies to the expression in Eq.~\eqref{eq:twoder}.} 
The term without any pion fields is obtained by the replacement $U\to \id, U^\dagger \to \id$, resulting in
\begin{align}
\label{eq:zeropi}
    F^2 B (m_u+m_d).
\end{align}
The proportionality constant $B$ is not constrained by chiral symmetry. 
But Eq.~\eqref{eq:zeropi} can be used to show that it is related to the vacuum expectation value $\langle \bar{q} q \rangle$, the so-called scalar quark condensate \citep{Gell-Mann:1968hlm,Gasser:1983yg}.

The term with two fields is given by
\begin{align}
    -\frac{1}{2} B (m_u + m_d) \sum_{a=1}^3 \phi_a \phi_a = -B(m_u + m_d)\pi^+ \pi^- -\frac{1}{2} B (m_u + m_d) \pi^0 \pi^0,
\end{align}
where the right-hand side follows from using Eq.~\eqref{eq:PiPhysCart} to rewrite the Cartesian pion fields in terms of the physical ones.
This corresponds to the mass terms of the pion fields. 
By comparing to the standard form of the mass term of charged and neutral pion fields,
\begin{align*}
    \calL_{M,\rm{charged}} &= - M^2 \pi^+ \pi^-  ,\\
    \calL_{M,\rm{neutral}} &=  - \frac{1}{2}M^2 \pi^0 \pi^0  ,
\end{align*}
 we can read off how the masses of the Goldstone bosons are related to the underlying quark masses, 
\begin{align}
\label{eq:LOPiMass}
    M^2 = B (m_u+m_d),
\end{align}
i.e., the square of the pion mass is proportional to the quark masses. 
Equation \eqref{eq:LOPiMass} represents the LO contribution to the square of the pion mass. 
The physical pion mass squared $M_\pi^2$ receives additional contributions at higher orders in the quark-mass expansion \citep{Gasser:1983yg}.

The term with four pion fields is given by
\begin{align}
    \frac{M^2}{24 F^2} \sum_{a,b=1}^3 \phi_a\phi_a \phi_b \phi_b.
\end{align}
It generates a contribution to the vertex in Fig.~\ref{fig:PiPiLO}. The combination of this term and the one of Eq.~\eqref{eq:twodertwopi}
generates the LO contributions to $\pi \pi$ scattering. 
There are two isospin channels, and the $S$-wave scattering lengths at LO are given by \citep{Weinberg:1966kf}
\begin{align}
\label{eq:pipiscatter}
    a_0^0 = \frac{7 M^2}{32 \pi F^2},\quad a_0^2 = - \frac{M^2}{16 \pi F^2}.
\end{align}
For a more detailed discussion, see, e.g.,~\citep{Meissner:2024ona}.
At this order, the scattering lengths depend only on the two LECs $F$ and $B$ (through $M^2$).
This demonstrates one of the advantages of the EFT approach. 
Each term in the Lagrangian contains an LEC, but because the Lagrangian is constructed in terms of $U$, the same LEC contributes to a variety of different processes in a predictable way.
Because $F$ can be determined from pion decay and $B$ is related to the pion mass, the result of Eq.~\eqref{eq:pipiscatter} is a prediction of \chipt without any free parameters.

Equation~\eqref{eq:LOPiMass} also helps us determine how the two identified expansion parameters, the momentum and the quark masses (divided by the scale $\Lambda$) are related to each other. 
For bookkeeping purposes, we introduce a generic small dimensionless parameter $q$ (not to be confused with a quark field), and each derivative acting on a pion field in the Lagrangian is of order $\calO(q)$.
For on-shell pions $p^2 = M_\pi^2$, i.e., $M_\pi^2\sim \calO(q^2)$. 
Equation~\eqref{eq:LOPiMass} then implies that the quark masses $\calM\sim \calO(q^2)$ as well.
In turn, this means that the Lagrangians of Eqs.~\eqref{eq:twoder} and \eqref{eq:Lsb} give contributions at the same order in the small-parameter expansion.
The leading-order effective Lagrangian is thus given by
\begin{align}
\label{eq:LOpiLag}
    \calL_2 = \frac{F^2}{4}\Tr(\partial_\mu U \partial^\mu U^\dagger)+\frac{F^2 B}{2} \Tr(\calM U^\dagger + U \calM^\dagger),
\end{align}
where the subscript 2 denotes that the Lagrangian generates contributions at order $\calO(q^2)$. 
This Lagrangian describes interactions between pions originating in QCD. 
But it does not include the fact that pions are also subject to electromagnetic and weak interactions. 
We now  discuss how to include these effects and the constraints imposed by the spontaneously broken chiral symmetry of QCD.

\subsection{Coupling to external fields and local chiral transformations}
\label{sec:local}

Instead of dealing with the electromagnetic and weak interactions of hadrons as full quantum field theories, their effects can be included by considering QCD coupled to external fields that represent these interactions.
To use the most general form, we add a Lagrangian $\Lext$ to the chiral-limit Lagrangian $ \calL_\text{QCD}^0$ of Eq.~\eqref{eq:chiralQCDLagLR}.
In the two-flavor case, $\Lext$ is given by 
\begin{align}
\label{eq:Lext}
    \Lext = \bar{q}_L \gamma^\mu \left( l_\mu +\frac{1}{3}v_\mu^{(s)} \right) q_L  
    + \bar{q}_R \gamma^\mu \left( r_\mu +\frac{1}{3}v_\mu^{(s)}\right) q_R 
    - \bar{q}_R \left(s+ip \right) q_L 
    - \bar{q}_L \left(s-ip \right) q_R .
\end{align}
With the exception of the so-called singlet-vector field $v_\mu^{(s)}$, the external fields are $2\times 2$ matrices given by
\begin{align}
    l_\mu = \sum_{a=1}^3 l_{\mu}^a\frac{\tau^a}{2}, \quad r_\mu = \sum_{a=1}^3 r_{\mu}^a \frac{\tau^a}{2}, \quad s = \sum_{a=0}^3 s^a\frac{\tau^a}{2}, \quad p = \sum_{a=0}^3 p^a \frac{\tau^a}{2},
\end{align}
where we use the convention that $\tau^0 = \id$.
The quark mass Lagrangian of Eq.~\eqref{eq:QCDMassTerm} is included in this formalism, as can be seen by setting $s+ip =\calM, s-ip = \calM^\dagger$ and all other fields to zero.
The coupling to electromagnetic and weak fields is achieved by appropriate choices for the external fields.
For example, the coupling to an electromagnetic potential $A_\mu$ is obtained by setting 
\begin{align}
\label{eq:EMext}
    r_\mu = l_\mu = -e \frac{\tau_3}{2}A_\mu,\quad v_\mu^{(s)} = - \frac{e}{2}A_\mu.
\end{align}
The coupling of quarks to weak fields can also be obtained by a different choice of the external fields, see below.
But the fields in Eq.~\eqref{eq:Lext} are not restricted to represent only standard model fields. 
They can also be used to study the coupling of quarks to beyond-the-standard-model physics. 
In this context, the coupling to tensor fields (which do not appear in the standard model) can also be included in $\Lext$ \citep{Cata:2007ns}.

The Lagrangian of Eq.~\eqref{eq:Lext} generalizes to the three-flavor case. 
In addition to the quark fields $q_{R/L}$ now being three-component vectors, the external fields are $3\times 3$ matrices with the general form
\begin{align}
    X = \sum_a X^a\frac{\lambda^a}{2}\, ,
\end{align}
where the $\lambda_a$ are the Gell-Mann matrices, $\lambda^0 = \id$, and $a$ is summed from 1 through 8 for $r_\mu$ and $l_\mu$ and from 0 to 8 for $s$ and $p$. The singlet-vector field $v_\mu^{(s)}$ is sometimes omitted in the three-flavor case because, unlike in the two-flavor case, it is not needed to reproduce the coupling to electromagnetic and weak fields.

The Lagrangian $\Lext$ breaks chiral symmetry explicitly. 
To map the coupling to external fields and the pattern of chiral symmetry breaking onto a \chipt Lagrangian, we follow the example of the quark mass term and determine how the external fields would have to transform to make $\Lext$ invariant.
In addition, we promote chiral symmetry from a global to a \emph{local} symmetry, i.e., the transformation matrices can now depend on $x$.
This is a theoretical tool -- the approximate chiral symmetry of QCD is a global symmetry, but considering it to be a local symmetry ensures that the effective Lagrangian reproduces the results of QCD; see~\citep{Gasser:1983yg,Gasser:1984gg} for more details.
Under local $\SU{2} \times \SU{2}$ transformations, the matrix $U(x)$ transforms as
\begin{align}
    U(x) \mapsto V_R(x) U(x) V_L(x)^\dagger.
\end{align}
In the following we will often suppress the explicit $x$ dependence, and the notation $V_L,V_R$ is used to distinguish the local $x$-dependent matrices from the global matrices $L,R$.
To maintain invariance of the Lagrangian under local transformations, the standard derivatives on the fields $U$ have to be replaced by covariant derivatives,
\begin{align}
\label{eq:picovder}
    D_\mu U \equiv \partial_\mu U -i r_\mu U + i U l_\mu,
\end{align}
which transform in the same way us $U$ itself, i.e,
\begin{align}
    D_\mu U \mapsto V_R (D_\mu U) V_L^\dagger.
\end{align}
At the same time, the external fields have to transform under local chiral $\SU{2} \times \SU{2}$ transformations as well,\footnote{In this formalism, the $\text{U}(1)_V$ symmetry is also promoted to a local symmetry, and $v_\mu^{(s)}$ transforms nontrivially as $v_\mu^{(s)} \to v_\mu^{(s)} - i \partial_\mu \Theta_V(x)$, where $\Theta(x)_V$ is the local analog of the phase in Eq.~\eqref{eq:Btransform}.}
\begin{align}
    s+ip & \mapsto V_R(s+ip)V_L^\dagger , &
    s-ip & \mapsto V_L(s-ip)V_R^\dagger , &
    r_\mu & \mapsto V_R r_\mu V_R^\dagger + i V_R \partial_\mu V_R^\dagger,&
    l_\mu & \mapsto V_L l_\mu V_L^\dagger + i V_L \partial_\mu V_L^\dagger, &
    v_\mu^{(s)} & \mapsto v_\mu^{(s)}.
\end{align}
Following the convention of~\citet{Gasser:1983yg}, we introduce
\begin{align}
    \chi = 2 B (s+ip).
\end{align}
With the replacement $s+ip \to\calM$, we see that it is related to $M^2$, and is therefore also of order $\calO(q^2)$ in terms of the generic small parameter $q$.
Because the fields $r_\mu$ and $l_\mu$ appear alongside the usual derivative $\partial_\mu$ in the covariant derivative of Eq.~\eqref{eq:picovder}, and the derivative on the pion fields is of order $\calO(q)$, $r_\mu$ and $l_\mu$ are also assigned to be of order $\calO(q)$, as is $v_\mu^{(s)}$.

Imposing invariance under local chiral transformations, the LO effective Lagrangian is obtained from Eq.~\eqref{eq:LOpiLag} by replacing the standard derivatives on $U$ with covariant derivatives and the quark mass matrix $\calM$ by $s+ip$ (expressed in terms of $\chi$). 
It thus takes the form
\begin{align}
\label{eq:L2}
    \calL_2 = \frac{F^2}{4}\Tr\left[ D_\mu U (D^\mu U)^\dagger\right]+\frac{F^2 }{4} \Tr(\chi U^\dagger + U \chi^\dagger).
\end{align}

We can now illustrate the earlier claim that the LEC $F$ is the pion decay constant. 
For concreteness, consider the $\pi^+$, with the valence quark structure $u\bar{d}$.
It can decay via the weak interaction, primarily into a muon and muon-neutrino, $\pi^+ \to \mu^+ + \nu_\mu$.
This decay is mediated via an intermediate $W^+$ boson, so we need to consider the coupling to an external field that reproduces this weak coupling.
This is obtained by setting all external fields to zero except $l_\mu$, for which we choose
\begin{align}
\label{eq:lweak}
    l_\mu = \sum_{a=1}^3 l_{\mu}^a\frac{\tau^a}{2} = -\frac{gV_{ud}}{\sqrt{2}} \left( W_\mu^+ \tau_+ + W_\mu^- \tau_- \right),
\end{align}
where $g$ is the weak gauge coupling, $V_{ud}$ the $ud$ element of the Cabibbo-Kobayashi-Maskawa quark mixing matrix (a real number), $W^{\pm}$ denotes the $W$ boson fields with $W^- = {(W^+)}^\dagger$, and $\tau_\pm=\frac{1}{2}(\tau_1\pm i \tau_2)$ with the $\tau_i$ the Pauli matrices acting in isospin space.
Pion decay is described by the Lagrangian in Eq.~\eqref{eq:L2} by terms that contain one pion field and one factor of $l_\mu$ given in Eq.~\eqref{eq:lweak}, which enters through the covariant derivative of Eq.~\eqref{eq:picovder}. 
With the fields $r_\mu$ set to zero, the covariant derivative simplifies to $\partial_\mu U+iUl_\mu$. 
Since we only need the terms with a single pion field, and the expansion of the derivative of $U$ contains at least one pion field, $\partial_\mu U = i \frac{\partial_\mu \phi}{F}+\cdots$, we can set $U=\id$ in the term with $l_\mu$, so that the covariant derivative can be replaced by $D_\mu \phi \to i \frac{\partial_\mu \phi }{F} + il_\mu$.
The relevant term in the Lagrangian is then given by
\begin{align}
    \calL_{2} & = \frac{F}{2}\Tr \left[\partial^\mu \phi l_\mu \right] +\cdots = \sum_{a=1}^3 \frac{F}{2} \partial^\mu \phi_a l_\mu^a +\cdots,
\end{align}
where the $l_{\mu}^a$ are determined from Eq.~\eqref{eq:lweak}.
The calculation of the pion-decay amplitude includes the hadronic matrix element 
\begin{align}
    \frac{F}{2} \langle 0 \vert \partial^\mu \phi_a \vert \phi_b \rangle = -i p^\mu \frac{F}{2} \delta_{ab},
\end{align}
which defines the pion decay constant.
The LEC $F$ can therefore be obtained by considering pion decay. For more details on how to perform the complete calculation, see, e.g.,~\citep{Scherer:2012xha}.

\subsection{Higher orders and power counting}
\label{sec:mesonpowercount}

The discussion so far has focused on terms that contribute to the effective Lagrangian at LO in an expansion in $q$.
Going beyond leading order requires additional terms in the effective Lagrangian.
These terms are constructed from the fields and their assigned order in the small expansion parameter identified above,
\begin{align}
    U\sim\calO(q^0), \quad D_\mu U \sim\calO(q^1), \quad r_\mu,l_\mu,v_\mu^{(s)} \sim\calO(q^1),  \quad \chi\sim\calO(q^2),
\end{align}
together with their transformation behavior under local chiral transformations given in Sec.~\ref{sec:local}.
The result is an expansion of the effective Lagrangian,
\begin{align}
    \calL_\text{eff} = \calL_2 + \calL_4 + \calL_6 +\ldots
\end{align}

But the goal is to calculate observables, which in a perturbative field theory can be obtained from amplitudes that are calculated through the use of Feynman diagrams. 
So we need a method of determining which diagrams contribute at a given order in an expansion in the small parameter $q$.
For mesonic \chipt, this method is Weinberg's power counting, which assigns a chiral dimension $D$ to each Feynman diagram obtained from $\calL_\text{eff}$.
To determine $D$, consider the amplitude $\mathscr{M}(p_i,m_q)$ of some diagram, depending on external momenta $p_i$ and the quark masses $m_q$, and simultaneously rescale the external momenta as $p_i \to t p_i$ and the quark masses as $m_q \to t^2 m_q$ (which due to Eq.~\eqref{eq:LOPiMass} corresponds to $M^2\to t^2 M^2$).
While external momenta can be changed in experiments, the quark masses are fixed in nature. 
But choosing $0<t< 1$, the rescaling of the quark masses can be thought of as a mathematical tool of approaching the chiral limit at the same time as considering low momenta.
Under these rescalings, the amplitude behaves as \citep{Weinberg:1978kz}
\begin{align}
\label{eq:AmpRescale}
    \mathscr{M}(t p_i,t^2 m_q) \to t^D \mathscr{M}(p_i,m_q).
\end{align}
Diagrams involving loops may be divergent and require regularization and renormalization.
In mesonic \chipt one typically uses dimensional regularization \citep{tHooft:1972tcz,tHooft:1978jhc,Leibbrandt:1975dj} in combination with a modified minimal subtraction scheme \citep{tHooft:1973mfk,Gasser:1983yg} (denoted in the following by $\widetilde{\text{MS}}$).
In dimensional regularization, the number of spacetime dimensions is generalized from 4 to $n$.
For diagrams with $N_{k}$ vertices from $\calL_{k}$, $N_I$ internal lines, and $N_L$ independent loops evaluated in $n$ spacetime dimensions, the chiral dimension is given by
\begin{align}
    D &=n N_L -2 N_I + \sum_{k\text{ even}}^\infty k\, N_{k}\\
    & = 2 + (n-2)N_L + \sum_{k \text{ even}}^\infty (k-2)\, N_{k}\\
    & \ge 2 \quad \text{for $n$=4 spacetime dimensions,}
\end{align}
where the equality of the first two lines follows from a relationship between the number of loops, internal lines, and the total number of vertices, $N_L = N_I-\sum N_k+1$.
For small $t$, Eq.~\eqref{eq:AmpRescale} shows that diagrams with small values of $D$ should be most important, while those with larger values of $D$ are suppressed.
Because $k\ge 2$, only a finite number of diagrams contribute to a given $D$.
This means that if we want to calculate an amplitude to order $q^N$, we need to consider all diagrams contributing to this amplitude for which $D\le N$.

\begin{figure}[t]
\centering
\begin{subfigure}[b]{0.3\textwidth}
\centering
    \includegraphics[width=.5\textwidth]{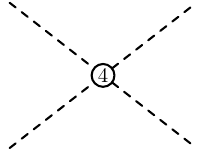}
    \caption{}\label{fig:PiPiNLO}
\end{subfigure}
\begin{subfigure}[b]{0.3\textwidth}
\centering
    \includegraphics[width=.63\textwidth]{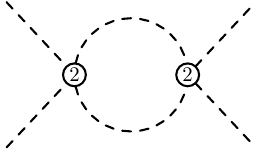}
    \caption{}\label{fig:PiPiLoop}
\end{subfigure}
\caption{Diagrams contributing to $\pi\pi$ scattering at chiral order $D=4$.}
\end{figure}

As an example, consider $\pi\pi$ scattering again.
We have already seen the diagram contributing at LO, see Fig.~\ref{fig:PiPiLO}. It has no loops ($N_L=0$), no internal lines ($N_I=0$), and a single vertex from $\calL_2$ ($k=2,N_2=1$), i.e., it corresponds to $D=2$.
The next-order contributions have $D=4$, shown in Figs.~\ref{fig:PiPiNLO} and \ref{fig:PiPiLoop}.
These consist of diagrams with $N_L=0,N_I=0,k=4,N_4=1$, i.e., tree-level diagrams with vertices from $\calL_4$, and diagrams with $N_L=1, N_I=2, k=2, N_2=2$, which are one-loop diagrams with two internal pion propagators and vertices from $\calL_2$.

\subsection{Generalization to three flavors}

The $s$ quark can also be incorporated in the chiral perturbation theory framework. For the three-flavor case, there are 8 Goldstone bosons: the $\pi$, $K$, and $\eta$  mesons. 
These are still collected in an exponential matrix $U(x)$ as in Eq.~\eqref{eq:Udef2}, which is now an \SU{3} matrix, with the Goldstone boson matrix taking the form
\begin{align}
    \phi & 
    =
    \begin{pmatrix}
        \pi^0 +\frac{1}{\sqrt{3}} \eta & \sqrt{2} \pi^+ & \sqrt{2} K^+\\
        \sqrt{2} \pi^- & -\pi^0 +\frac{1}{\sqrt{3}} \eta & \sqrt{2} K^0 \\
        \sqrt{2} K^- & \sqrt{2} \bar{K}^0 & -\frac{2}{\sqrt{3}} \eta  
    \end{pmatrix}.
\end{align}
All external fields are also generalized to their \SU{3} versions. 
The form of the LO Lagrangian is completely analogous to Eq.~\eqref{eq:L2}, the only difference is that the \SU{2} LECs $F, B$ are replaced by their \SU{3} analogs $F_0, B_0$ (with $B_0$ hidden in the definition of $\chi$),
\begin{align}
    \calL_2 = \frac{F_0^2}{4}\Tr(D_\mu U D^\mu U^\dagger)+\frac{F_0^2 }{4} \Tr(\chi U^\dagger + U \chi^\dagger).
\end{align}
While the power counting also generalizes straightforwardly, at higher orders the form of the effective Lagrangian will be different from the \SU{2} case. 
This stems from the application of certain trace relations in reducing the Lagrangian to a minimal form, and these trace relations are different for \SU{2} matrices compared to \SU{3} matrices.
The convergence in the \SU{2} case is expected to be better than for \SU{3}. 
The $K$ and $\eta$ mesons are significantly heavier than the pions, so that an expansion in $M_K^2/\Lambda^2$ and $M_\eta^2/\Lambda^2$ converges more slowly than one in $M_\pi^2/\Lambda^2$.
Nonetheless, \SU{3} \chipt serves as a well-defined EFT to describe low-energy strong interactions including strange particles.

\section{Baryon chiral perturbation theory}
\label{sec:bchpt}

\subsection{The LO Lagrangian}

Low-energy nuclear physics typically involves protons and neutrons.
How can these be incorporated into the \chipt framework?
Since protons and neutrons are not Goldstone bosons, a different description than the one used for pions is required.
As discussed in Sec.~\ref{sec:isospin}, the proton $p$ and neutron $n$ are considered as two states of a nucleon, which is described as an isospin-1/2 field $\Psi$,\footnote{This notation, which differs from the one used in Sec.~\ref{sec:isospin}, is the standard notation in baryon \chipt. The fields $p$ and $n$ in $\Psi$ are themselves four-component Dirac spinors, which is irrelevant for the discussion of Sec.~\ref{sec:isospin}.}
\begin{align}
    \Psi = 
    \begin{pmatrix}
        p \\
        n
    \end{pmatrix}.
\end{align}
Under isospin transformations $V\in \SU{2}$, the nucleon field should transform linearly as $\Psi \mapsto V \Psi$.
Under chiral transformations $(L,R)\in \SU{2}_L \times \SU{2}_R$, we choose $\Psi$ to transform as \citep{Georgi:WeakInt}
\begin{align}
    \Psi \mapsto K(L,R,U) \Psi,
\end{align}
where the \SU{2} matrix $K(L,R,U)$ is given by
\begin{align}
\label{eq:Kdefine}
    K(L,R,U) \equiv {\sqrt{RUL^\dagger}}^{\,-1} R \sqrt{U}.
\end{align}
As expected, this transformation depends on $L$ and $R$. 
But it also depends on the pion fields, encoded in $U$, and since $U$ is a function of $x$, $K(L,R,U)$ is local even for global chiral transformations $(L,R)$.
An isospin transformation corresponds to $L=R=V$, and in this case $K(V,V,U)=V$ (see, e.g., the discussion in~\citep{Scherer:2012xha}). 
In this special case, the transformation of the nucleon is independent of $U$, and $\Psi$ transforms linearly as required.
While $K(L,R,U)$ looks very different from the transformation of the pion fields $U$ in Eq.~\eqref{eq:Utransform}, we could have also described meson \chipt in terms of $K$.
Define the unitary square root $u$ of $U$,
\begin{align}
    [u(x)]^2 = U(x).
\end{align}
It transforms under chiral transformations as
\begin{align}
    u\mapsto \sqrt{RUL^\dagger} = R u K^{-1}(L,R,U),
\end{align}
where the last equality can be shown using the definition of Eq.~\eqref{eq:Kdefine}.
However, for mesonic \chipt the formalism of Sec.~\ref{sec:mesonchpt} is typically more convenient.
The exponential form of $U$ given in Eq.~\eqref{eq:Udef2} is particularly convenient, since then $u$ takes the form
\begin{align}
\label{eq:udef}
    u(x)= \exp\left[i\frac{\phi(x)}{2F} \right] = \id + i\frac{\phi(x)}{2F} -  \frac{\phi^2}{8F^2}+\cdots.
\end{align}

For processes involving one nucleon in the initial state and one nucleon in the final state, the effective Lagrangian contains terms of the form
\begin{align}
    \bar{\Psi} \hat{O} \Psi,
\end{align}
where $\hat{O}$ denotes an operator built out of derivatives, pions, and the external fields.
For the Lagrangian to be invariant under chiral transformations, the operators must transform as $K\hat{O}K^\dagger$.
Like in the meson case, we promote chiral symmetry from a global symmetry to a local symmetry, i.e., we consider $K=K(V_L(x),V_R(x),U(x))$.
For these local transformations, the standard derivative acting on the nucleon field is replaced by a covariant derivative,
\begin{align}
    D_\mu \Psi = \left( \partial_\mu +\Gamma_\mu -iv_\mu^{(s)} \right) \Psi,
\end{align}
where the so-called chiral connection is defined as
\begin{align}
\label{eq:chiconnect}
    \Gamma^\mu \equiv \frac{1}{2} \left[ u^\dagger (\partial_\mu -ir_\mu)u +u (\partial_\mu -il_\mu) u^\dagger \right].
\end{align}
The nucleon covariant derivative transforms like the nucleon field, i.e.,
\begin{align}
    D_\mu \Psi \mapsto K(V_L(x),V_R(x),U(x))\, D_\mu \Psi.
\end{align}
We also introduce the chiral vielbein,
\begin{align}
\label{eq:chiviel}
    u_\mu \equiv i \left[ u^\dagger (\partial_\mu -ir_\mu)u - u (\partial_\mu -il_\mu) u^\dagger \right],
\end{align}
which has the desired transformation behavior $u_\mu \mapsto K u_\mu K^\dagger$.
The most general nucleon Lagrangian invariant under local chiral transformations and the discrete symmetries with at most one derivative is given by
\begin{align}
\label{eq:LpiNLO}
    \calL_{\pi N}^{(1)} = \bar{\Psi} \left( i \slashed{D} - m + \frac{\gCL}{2} \gamma^\mu \gamma_5 u_\mu \right) \Psi.
\end{align}
The two parameters $m$ and $\gCL$ are the nucleon mass and the nucleon axial-vector coupling constant, respectively, both considered in the chiral limit.
These are different from their physical values, which are denoted by $m_N$ and $g_A$, respectively.
The nucleon axial-vector coupling $g_A$ can be determined from neutron $\beta$ decay, $n\to p + e^- + \bar{\nu}_e$.

\begin{figure}[t]
\centering
\begin{subfigure}[b]{0.3\textwidth}
\centering
    \includegraphics[width=.6\textwidth]{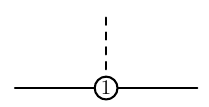}
    \caption{}\label{fig:NNpi}
\end{subfigure}
\begin{subfigure}[b]{0.3\textwidth}
\centering
    \includegraphics[width=.6\textwidth]{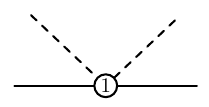}
    \caption{}\label{fig:NNpipi}
\end{subfigure}
\begin{subfigure}[b]{0.3\textwidth}
\centering
    \includegraphics[width=.6\textwidth]{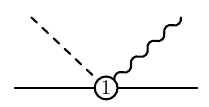}
    \caption{}\label{fig:NNpigamma}
\end{subfigure}
\caption{Vertices obtained from $\calL_{\pi N}^{(1)}$.}
\end{figure}

\begin{figure}[t]
\centering
\begin{subfigure}[b]{0.3\textwidth}
\centering
    \includegraphics[width=.6\textwidth]{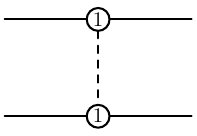}
    \caption{}\label{fig:OPE}
\end{subfigure}
\begin{subfigure}[b]{0.3\textwidth}
\centering
    \includegraphics[width=.6\textwidth]{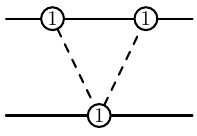}
    \caption{}\label{fig:TPE}
\end{subfigure}
\begin{subfigure}[b]{0.3\textwidth}
\centering
    \includegraphics[width=.6\textwidth]{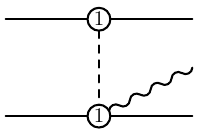}
    \caption{}\label{fig:MEC}
\end{subfigure}
\caption{Pion exchange contributions to \NN interactions and currents.}
\end{figure}

If we set all external fields to zero and and consider only the first term in the expansion of Eq.~\eqref{eq:udef}, i.e., the term without pion fields, $\calL_{\pi N}^{(1)}$ reduces to the Lagrangian of a free nucleon of mass $m$. 
But the general Lagrangian also describes the interactions of a nucleon with one, two, and more pions and/or external fields.
Let us consider a few examples. 
The term with a single pion field has the form
\begin{align}
    -\frac{\gCL}{2F} \bar{\Psi}\gamma^\mu \gamma_5 \tau^a \, \partial_\mu \pi_a\,  \Psi.
\end{align}
It describes a pion coupling to a nucleon, see the vertex in Fig.~\ref{fig:NNpi}.
This form of the interaction, with a derivative acting on the pion field and a fermion bilinear of the type $\bar{\Psi} \gamma^\mu \gamma_5 \Psi$, is a pseudo-vector pion-nucleon interaction.
It forms the basis of the one-pion-exchange contribution to the nucleon-nucleon (\NN) interactions, see Fig~\ref{fig:OPE}. 
However, for \NN interactions one typically uses a nonrelativistic reduction of these expressions, see the discussion in Sec.~\ref{sec:BChPTRenorm} below.
The term with two pion fields (and still no external fields) is given by
\begin{align}
    -\frac{1}{4F^2} \epsilon_{abc} \bar{\Psi} \gamma^\mu \pi^a \partial_\mu \pi^b \tau^c \Psi.
\end{align}
It describes two pions interacting with the nucleon at the same vertex, see Fig.~\ref{fig:NNpipi}. 
It contributes, e.g., to $\pi N$ scattering, but also to two-nucleon exchange diagrams of the \NN interactions, see Fig~\ref{fig:TPE}.
As the final example, consider the term with a single pion field and the coupling to an external electromagnetic field as given in Eq.~\eqref{eq:EMext}.
It results in a term of the form
\begin{align}
    -e \frac{\gCL}{2F} \epsilon_{3ab} \bar{\Psi} \gamma^\mu \gamma_5 \tau_a\, A_\mu \pi^b \Psi .
\end{align}
This is the so-called Kroll-Ruderman term, representing the interaction of a nucleon with a pion and a photon at the same vertex, see Fig.~~\ref{fig:NNpigamma}.  
It contributes, e.g., to pion photoproduction on the nucleon, $\gamma + N \to \pi + N$.
It also contributes to the \NN meson-exchange currents, in which a photon couples to a nucleon, which in turn exchanges a pion with the second nucleon, see Fig.~\ref{fig:MEC}.
For a more detailed discussion of how baryon \chipt contributes to \NN interactions within an EFT framework, see~\citep{EncycNN} and \citep{Hammer:2019poc,Machleidt:2024bwl} for some recent reviews. 

Because other choices for the external fields are possible and the matrix $u$ contains an infinite tower of terms with an increasing number of pion fields, these are just a few examples of the interactions described by the Lagrangian $\calL_{\pi N}^{(1)}$. 
All of these interactions depend on some combination of the two LECs $F$ (the pion decay constant in the chiral limit) and $\gCL$ (the nucleon axial-vector coupling in the chiral limit).

\subsection{Power counting for baryon \chipt}

For the Lagrangian of Eq.~\eqref{eq:LpiNLO}, we restricted the discussion to terms with at most one derivative.
As discussed in Sec.~\ref{sec:mesonchpt}, for pions each derivative corresponds to an additional factor of the small parameter $q$, and the mesonic Lagrangian can be expanded in terms with an increasing number of derivatives.
But the analogous argument does not hold for nucleons. 
The zeroth component of the derivative corresponds to the nucleon energy.
Even for nucleons with small three-momenta of order $\calO(q)$, this energy is at least $m$, the nucleon mass in the chiral limit.
But since nucleons are not Goldstone bosons, there is no reason why their mass should vanish or become small (compared to $\Lambda$) in the chiral limit.
In fact, baryon \chipt has been used to determine $m\approx \SI{870}{MeV}\text{--}\SI{880}{MeV}$ \citep{Fuchs:2003kq,Hoferichter:2015hva}, which is at least of the same order as the scale $\Lambda$. 
Instead, the term $(i\slashed{D}-m)\Psi$ is counted as a term of order $\calO(q)$, since the large contribution from the nucleon mass cancels.

Further, the Lagrangian consists of bilinears of the form $\bar{\Psi} \Gamma \Psi$, where $\Gamma$ denotes the possible combinations of Dirac matrices.
For nonrelativistic nucleons, the Dirac spinors can be thought of as consisting of a large upper component and a suppressed lower component. 
In the bilinears, some Dirac matrices couple large with large components, while others only couple large with small components. 
One therefore also assigns powers of $q$ to the Dirac matrices.
In summary, the new elements that appear in baryon \chipt follow the power counting:
\begin{align}
    \bar{\Psi}, \Psi \sim \calO(q^0), \quad D_\mu \Psi \sim \calO(q^0) \quad (i\slashed D - m)\Psi \sim \calO(q^1), \quad \id,\gamma_\mu,\gamma_\mu\gamma_5,\sigma_{\mu\nu}\sim \calO(q^0), \quad \gamma_5 \sim \calO(q^1).
\end{align}
Higher-order terms in the Lagrangian can then be constructed, see, e.g.,~\citep{Gasser:1987rb,Fettes:2000gb} for details.
Unlike in the mesonic case, there are terms at each order in the $q$ expansion, not just the even order, i.e.,
\begin{align}
    \calL_{\pi N} = \calL_{\pi N}^{(1)} + \calL_{\pi N}^{(2)} + \calL_{\pi N}^{(3)} + \calL_{\pi N}^{(4)} +\cdots.
\end{align}
For example, the next-to-leading order Lagrangian $\calL_{\pi N}^{(2)}$ contains seven independent terms, each with a new LEC, typically denoted $c_i$.
Some of these are important in $\pi N$ scattering and also contribute to pion-exchange terms in the \NN interactions.

In addition to the Lagrangian, we also need power counting rules for diagrams. 
For mesons, the rules of Sec.~\ref{sec:mesonpowercount} still hold.
In addition, vertices from the $\pi N$ Lagrangian at order $k$ count as $q^k$.  
A nucleon propagator is proportional to $(\slashed{p} - m)^{-1}$, and so the counting of $(i\slashed D - m)\Psi$ suggests that it should be counted as $q^{-1}$.
The chiral dimension $D$ of a diagram involving nucleons is therefore expected to be given by
\begin{align}
\label{eq:baryonpowercount}
    D=n N_L -2 I_\pi -I_N + \sum_{k\text{ even}}^\infty k\, N_{k}^{(\pi)} + \sum_{k'}^\infty k'\, N_{k'}^{(\pi N)},
\end{align}
where $N_L$ is again the number of independent loops, $I_\pi$ and $I_N$ are the number of internal pion and nucleon lines, and $N_k^\pi$ and $N_{k'}^{\pi N}$ are the number of vertices derived from the mesonic Lagrangian $\calL_k$ and the $\pi N$ Lagrangian $\calL_{\pi N}^{(k')}$, respectively.

\begin{figure}[t]
\centering
    \includegraphics[width=.2\textwidth]{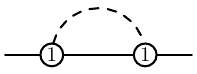}
    \caption{One-loop contribution to the nucleon self energy}\label{fig:Nmass}
\end{figure}

However, the results of loop diagrams in general do not agree with the power counting rule of Eq.~\eqref{eq:baryonpowercount}. 
For example, the loop diagram of Fig.~\ref{fig:Nmass}, which contributes to the nucleon mass, contains one loop ($N_L=1$), one internal pion line ($I_\pi=1$), one internal nucleon line ($I_N=1$), and two vertices from $\calL_{\pi N}^{(1)}$ ($k'=1, N_{1}^{(\pi N)}=2$).
According  to Eq.~\eqref{eq:baryonpowercount} its chiral dimension should therefore be $D=4-2-1+2=3$ in $n=4$ spacetime dimensions.
But the expression obtained by applying the same methods of dimensional regularization and the \MS renormalization scheme of the meson sector (see the discussion below Eq.~\eqref{eq:AmpRescale}) contains a term of order $\calO(q^2)$, i.e., lower than the expected $D=3$ \citep{Gasser:1987rb}.
This can be traced back to the fact that the nucleon mass does not vanish in the chiral limit. As \citet{Gasser:1987rb} state: ``This complicates life a lot.''
There are different resolutions to this power counting puzzle, which we discuss next. 
Ultimately, this issue is related to regularization and renormalization, and \citet{Gasser:1987rb} also point out that the ``same phenomenon would occur in the meson sector, if one did not make use of dimensional regularization.''

\subsection{Resolving the power counting puzzle}
\label{sec:BChPTRenorm}

There are several ways to address the power counting issue.
The three most prominent and most widely-used ones are the heavy-baryon approach \citep{Jenkins:1990jv,Bernard:1992qa}, the method of infrared regularization (IR) \citep{Becher:1999he}, and the extended-on-mass-shell (EOMS) scheme \citep{Gegelia:1999gf,Fuchs:2003qc}.
They have in common that their application results in expressions for loop diagrams that satisfy the power counting of Eq.~\eqref{eq:baryonpowercount}.
The heavy-baryon approach involves an additional expansion of the Lagrangian in powers of $1/m$ and is thus similar to a nonrelativistic expansion, while both the IR and the EOMS schemes maintain manifest Lorentz invariance.
Historically the heavy-baryon approach was the first of these three methods, and it is the one that is typically used when the chiral EFT approach is extended to the two- and few-nucleon sectors.
However, we begin with a brief discussion of the IR and EOMS schemes, which share some commonalities. 

Infrared regularization is based on the analysis of infrared singularities that arise for small momenta.
A one-loop diagram results in an integral over the loop momentum, which can be translated into an integral over a Feynman parameter of the type
\begin{align}
\label{eq:Hdef}
    H=\int_0^1 dz\, C(z),
\end{align}
where $C(z)$ can depend on external momenta, the pion mass, and nucleon mass.
In IR regularization this integral is split into two pieces,
\begin{align}
    H &= I+R \\
    & = \int_0^\infty dz\, C(z) - \int_1^\infty dz\, C(z).
\end{align}
The crucial observation is that the terms that violate the power counting are contained in the so-called infrared-regular term $R$, that $R$ is analytic in small quantities, and that it is chirally invariant by itself.
This means that $R$ can be expanded in these small quantities (we will refer to this as the chiral expansion of the expression) and each term can be absorbed by the LECs of the Lagrangian.
The remaining infrared-singular part $I$ only contains terms that satisfy the power counting, and it is also chirally invariant. 
The IR method consists of replacing each integral $H$ by its infrared-singular part $I$, which maintains chiral symmetry and results in expressions that satisfy the power counting rule of Eq.~\eqref{eq:baryonpowercount}.

The EOMS scheme uses a somewhat similar approach of using a renormalization prescription to absorb the power-counting-violating terms.
In the IR approach, these terms are contained in the infrared-regular part $R$.
But the chiral expansion of $R$ can also contain terms that satisfy the power counting.
By removing $R$ completely, the IR approach absorbs both types of terms in the LECs.
In contrast, in the EOMS scheme only the terms that violate the power counting are removed by renormalization.
These terms can be identified by expanding the integrand of a loop integral $H$ in small quantities \emph{before} integration, and then subtracting from the integrand the power-counting-violating terms. 
The remainder is the renormalized expression in the EOMS scheme,
\begin{align}
    H_R=H-H_\text{subtr},
\end{align}
and it satisfies the power counting.

In low-energy nuclear physics, nucleons are frequently treated as nonrelativistic particles, with three-momenta much smaller than the nucleon mass $m$.
Heavy-baryon \chipt (HB\chipt) provides a framework for a nonrelativistic description of nucleons interacting with pions and external fields that also solves the power counting issue.
It typically serves as the basis for constructing \NN interactions within a chiral EFT framework.
The basic idea behind HB\chipt is to perform an additional expansion of the baryon Lagrangian in powers of $1/m$.
It shares similarities with the Foldy-Wouthuysen expansion of the Dirac equation \citep{Foldy:1949wa} and with heavy quark effective theory, an EFT for describing baryons that contain a heavy $b$, $c$, or $t$ quark \citep{Isgur:1989vq,Eichten:1989zv,Georgi:1990um} (also see, e.g, the review in~\citep{PDG} and~\citep{EncycPartHQET} for an introduction).
To perform the additional expansion, the nucleon four-momentum is separated into a large part that is close to the mass shell, and a small (soft) residual piece,
\begin{align}
    p^\mu = m v^\mu + k^\mu_p.
\end{align}
The four-vector $v^\mu$ satisfies
\begin{align}
    v^2 = 1, \quad v^0\ge 1, \quad v\cdot k_p \ll m.
\end{align}
For convenience, it is often taken to be $v^\mu = (1,0,0,0)$.
The nucleon field is separated into two fields,
\begin{align}
    \Psi(x) = e^{-imv\cdot x} \left[ \calN_v(x) + \calH_v(x) \right],
\end{align}
where $\calN_v$ is called the light component amd $\calH_v$ the heavy component.
Defining projection operators 
\begin{align}
    P_{v\pm} \equiv \frac{1}{2} (1\pm \slashed{v}),
\end{align}
the light and heavy components are given by
\begin{align}
    \calN_v = e^{imv\cdot x} P_{v+}\Psi, \quad \calH_v = e^{imv\cdot x} P_{v-}\Psi.
\end{align}
By applying the projection operators $P_{v\pm}$ onto the equation of motion (EoM) for the nucleon field $\Psi$, one obtains two separate EoMs the fields $\calN_v$ and $\calH_v$ must satisfy.
These can be used to eliminate the heavy component, resulting in an EoM for the light component $\calN_v$. 
This EoM is also obtained from the Lagrangian
\begin{align}
\label{eq:HBLag}
    \widehat{\calL}_{\text{eff}} & = \widehat{\calL}_{\pi N}^{\,(1)} + O(1/m)\\
    & = \bar{\calN}_v \left( i v\cdot D + \gCL S_{\! v} \cdot u \right) \calN_v + O(1/m).
\end{align}
The symbol $\widehat{\phantom{\calL}}$ is used to denote the heavy-baryon formalism, and the spin matrix $S_{\! v}$ is defined as
\begin{align}
    S^\mu_{\! v} \equiv \frac{i}{2}\gamma_5 \sigma^{\mu\nu} v_\nu . 
\end{align}
The term $O(1/m)$ in Eq.~\eqref{eq:HBLag} indicates that the effective Lagrangian that reproduces the EoM for $\calN_v$ has been expanded in inverse powers of the nucleon mass.
The full heavy-baryon Lagrangian also contains higher-order terms that do not originate from the $1/m$ expansion, analogous to those in the Lorentz-invariant case, see~\citep{Fettes:2000gb} for more details.

Let us return to the power counting issue.
The heavy-baryon Lagrangian at LO is independent of the nucleon mass $m$, i.e., the nucleon mass has does not appear at all in the LO piece $\widehat{\calL}_{\pi N}^{\,(1)}$.
The nucleon Feynman propagator derived from $\widehat{\calL}_{\pi N}^{\,(1)}$ therefore does not depend on $m$ either,
\begin{align}
\label{eq:HBprop}
    \mathcal{S}_{\! F}(k) = \frac{P_{v+}}{v\cdot k +i0^+}\, .
\end{align}
Because the nucleon mass can only enter as factors of $1/m$ from the expansion of the effective Lagrangian, diagrams computed with the heavy baryon propagator of Eq.~\eqref{eq:HBprop} and vertices derived from the heavy-baryon Lagrangian automatically satisfy the power counting of Eq.~\eqref{eq:baryonpowercount} when applying dimensional regularization on the \MS renormalization scheme just as in the mesonic sector.

\subsection{Generalization to three flavors}

Baryon \chipt can also be extended to include the $s$ quark.
Instead of the isospin doublet $N$, one considers the octet of $\frac{1}{2}^+$ baryons consisting of nucleons, $\Sigma$'s, $\Xi$'s, and the $\Lambda$ (the baryons containing one or more strange quarks are called hyperons). 
These are collected in a traceless $3\times 3$ matrix $B$,
\begin{align}
    B = 
    \begin{pmatrix}
        \frac{1}{\sqrt{2}} \Sigma^0 + \frac{1}{\sqrt{6}} \Lambda & \Sigma^+ & p \\
        \Sigma^- & -\frac{1}{\sqrt{2}} \Sigma^0 + \frac{1}{\sqrt{6}} \Lambda & n \\
        \Xi^- & \Xi^0 & -\frac{2}{\sqrt{6}} \Lambda
    \end{pmatrix}.
\end{align}
Under chiral transformations, the baryon matrix transforms as $KBK^\dagger$. 
The chiral connection $\Gamma_\mu$ of Eq.~\eqref{eq:chiconnect} and the chiral vielbein $u_\mu$ of Eq.~\eqref{eq:chiviel} generalize straightforwardly to the three-flavor case by using the \SU{3} versions of $u$ and the external fields.
The covariant derivative for $B$ is given by
\begin{align}
    D_\mu B =\partial_\mu B + \left[ \Gamma_\mu,B \right],
\end{align}
and the LO Lagrangian has the form
\begin{align}
    \calL_{\text{MB}}^{(1)} = \Tr\left[ \bar{B} (i\slashed{D}-M_0) B \right] 
    - \frac{\mathtt{D}}{2} \Tr\left[ \bar{B} \gamma^\mu \gamma_5 \{ u_\mu,B\} \right]
    - \frac{\mathtt{F}}{2} \Tr\left[\bar{B} \gamma^\mu \gamma_5 [ u_\mu,B] \right].
\end{align}
$M_0$ is the baryon octet mass in the chiral limit, and there are now two independent LECs $\mathtt{D}$ and $\mathtt{F}$ instead of the single LEC $\gCL$ in the \SU{2} sector.\footnote{We use the notation $\mathtt{F}$ for the baryon LEC to distinguish it from $F$, the \SU{2} version of the pion decay constant in the chiral limit.}
One can also perform a heavy-baryon expansion of the \SU{3} Lagrangian in analogy to the two-flavor case. 
The meson-baryon interaction vertices obtained from the Lagrangian can be used in constructing meson-exchange contributions to hyperon-nucleon (and in principle hyperon-hyperon) interactions; see, e.g.,~\citep{Petschauer:2020urh,Haidenbauer:2022esw} for recent reviews on the hyperon-nucleon interactions.

\section{The large-N$_\text{c}$ limit}
\label{sec:largeN}

Color $\SU{3}_c$ gauge symmetry is the defining feature of QCD. 
But as argued in Sec.~\ref{sec:QCDLag}, if we formulate an effective theory in terms of color-neutral hadrons, color gauge symmetry does not provide any direct constraints on the interactions between the hadrons.
Ultimately, these interactions are still governed by how quarks and gluons interact, but the nonperturbative nature of QCD at low energies makes it very difficult to establish a direct connection between QCD and the hadronic theories.
It turns out that QCD simplifies when the theory is generalized from 3 colors to an $\SU{\Nc}$ gauge theory, where \Nc stands for the number of colors, in the limit that \Nc is taken to infinity \citep{tHooft:1973alw}.
But even if the theory does become simpler (see, e.g., the reviews~\citep{Witten:1979kh,Coleman:1985rnk,Manohar:1998xv,Jenkins:1998wy,Lebed:1998st} for detailed discussions and an overview of results), how do results obtained in the large-\Nc limit relate to the physical world we live in, where $\Nc=3$?
The large-\Nc approach provides a method to not only consider physical quantities in the $\Nc \to \infty$ limit, but to also determine corrections in an expansion in $1/\Nc$.
It turns out that there are many cases for which these results agree with experimental observations when setting $\Nc=3$, even though $1/3$ is not a very small expansion parameter.

Under the change to an \SU{\Nc} gauge symmetry, many of the other (approximate) symmetries of QCD remain intact, since they are typically related to the flavor, rather than the color, aspects of QCD.
One difference emerges for the axial $\text{U}(1)_A$ transformations that leave the QCD Lagrangian invariant in the chiral limit. 
As discussed in Sec.~\ref{sec:anomaly}, in the physical world ($\Nc=3$), quantum effects break this symmetry.
However, these quantum effects are suppressed in the large-\Nc limit, so that $\text{U}(1)_A$ is a symmetry of QCD in the chiral and large-\Nc limits.
Chiral symmetry is still spontaneously broken \citep{Coleman:1980mx}, but the breaking pattern is now from $\text{U}(N)_L \times \text{U}(N)_R$ to $\text{U}(N)_V$, with $N=2$ or $3$ the number of light quark flavors. 
As a result, there is now an extra Goldstone boson.
In the three-flavor case, this is identified as the $\eta'$ meson.
One can then formulate a large-\Nc variant of \chipt, see, e.g.,~\citep{Moussallam:1994xp,Leutwyler:1996sa,Kaiser:2000gs}.

While mesons retain their quark-antiquark valence quark picture in the large-\Nc limit and many properties can be derived on the basis of Feynman diagram considerations, the situation is different for baryons.
With \Nc colors, baryons consist of \Nc valence quarks. 
While the analysis is more complex, many interesting results can still be derived, see~\citep{Jenkins:1998wy} for a review.
One of these is that the baryon mass grows with \Nc \citep{Witten:1979kh}. 
The second result of interest here is that in order to obtain a sensible large-\Nc limit for the baryon-meson scattering amplitude, an infinite tower of baryon states with the same mass, but increasing total spin $\mathcal{S}$ and isospin $I$ with $\mathcal{S}=I=\frac{1}{2},\frac{3}{2},\frac{5}{2},\ldots$ is required \citep{Gervais:1983wq, Gervais:1984rc,Dashen:1993as,Dashen:1993ac}.
Nucleons correspond to the $\mathcal{S}=I=\frac{1}{2}$ states, and the $\mathcal{S}=I=\frac{3}{2}$ states are identified with the $\Delta$ baryons.\footnote{States with higher $\mathcal{S}=I$ only exist for $\Nc>3$, but not in the physical world for which $\Nc=3$.}
In addition, there are relationships among the couplings of these baryons to mesons.
For example, the ratio of the coupling $g_{\pi \Delta N}$ of a $\Delta$ to a pion and a nucleon to the $\pi N N$ coupling $g_{\pi N N}$ is given by
\begin{align}
    \frac{g_{\pi \Delta N}}{g_{\pi N N}} = \frac{3}{2}.
\end{align}
These relationships are equivalent to a combined spin-flavor symmetry \SU{2N} among baryons \citep{Dashen:1993as,Dashen:1993ac,Dashen:1993jt}.\footnote{Technically, this is a contracted symmetry, see, e.g.,~\citep{Jenkins:1998wy}.}
Here, $N$ is the number of flavors, i.e., this symmetry is an \SU{4} symmetry when considering only $u$ and $d$ quarks, and an \SU{6} symmetry when including the $s$ quark as well.
This new symmetry that emerges for baryons in the large-\Nc limit is \emph{not} a symmetry of QCD in the large-\Nc limit and has no impact on the meson sector, but it provides constraints on baryon matrix elements.
For example, baryon masses, magnetic moments, and axial couplings have been analyzed in the large-\Nc framework, see the review by~\citet{Jenkins:1998wy} and references therein.

The large-\Nc constraints on the baryon matrix elements can also be used to  constrain interactions between two and more nucleons.
These constraints typically take the form of relationships between various parameters of the \NN interactions.
As an example, consider \NN interactions at low energies.
There are two independent terms for $S$-wave interactions: one for the \oneS channel and one for the \threeS channel, with the notation corresponding to ${}^{3\mathcal{S}+1}\!L_J$, where $\mathcal{S},L,J$ denote the total spin, orbital angular momentum, and total angular momentum of the \NN state, respectively.
In the large-\Nc limit, the interactions in these two channels become equal to each other, i.e., there is now only a single independent term that governs the interactions in the two channels \citep{Kaplan:1995yg,Kaplan:1996rk}.
As a result, the $S$-wave \NN interactions become invariant under an \SU{4} spin-isospin symmetry, called Wigner symmetry \citep{Wigner:1936dx}.
The nucleons are collected into a four component vector $(p\!\uparrow, p\!\downarrow,n\!\uparrow, n\!\downarrow)$, on which the \SU{4} transformations act.
While tensor interactions and interactions including higher partial waves are not Wigner-\SU{4}-invariant, the symmetry is still useful in nuclear physics, see, e.g.,~\citep{Vogel:1993zz}. 
Recent applications of Wigner-\SU{4} symmetry can be found in~\citep{Vanasse:2016umz,Lee:2020esp,Lin:2022yaf,Nguyen:2024rlr,LiMuli:2025zro}, for instance.

Large-\Nc methods have also been applied to \NN interactions beyond $S$-waves \citep{Schindler:2018irz}, to three-nucleon interactions \citep{Phillips:2013rsa,Epelbaum:2014sea}, and to the coupling to external currents \citep{Richardson:2020iqi}.
Analogously to the $S$-wave sector, the baryon-sector symmetry that emerges in the large-\Nc limit imposes constraints on these interactions, typically reducing the number of independent terms that need to be considered. 
For a review see~\citep{Richardson:2022hyj}.

\section{Conclusions}
\label{sec:conc}

Even though a direct description of low-energy nuclear physics in terms of QCD remains challenging, the symmetries and symmetry-breaking patterns of QCD are reflected in low-energy observables.
Any theory that aims to reproduce the results of QCD at low energies should possess the same exact and approximate symmetries. 
This requirement provides important constraints on effective descriptions of low-energy nuclear physics.
We have shown how the symmetries of QCD, in particular chiral symmetry and its spontaneous and explicit breaking, form the basis of \chipt, the low-energy effective field theory of QCD for pions and nucleons. 
\chipt has been extremely successful, and it serves as the basis of extending the EFT approach to interactions between two and more nucleons.
Thus, the symmetries of QCD form the foundation of modern descriptions of low-energy nuclear physics.

\begin{ack}[Acknowledgments]

I thank S.~Scherer for introducing me to the world of chiral dynamics, and I am grateful to all my collaborators for many interesting discussions that have have contributed to my understanding of these topics.
Special thanks to R.~P.~Springer for feedback on the manuscript.
This work was supported by the U.S.~Department of Energy, Office of Science, Office of Nuclear Physics, Award No.~DE-SC0019647.

\end{ack}

\seealso{The nuclear force: the first sixty years , EFT and renormalization, Nuclear forces from lattice QCD, Chiral EFT for nuclear forces}

\begin{thebibliography*}{110}
\providecommand{\bibtype}[1]{}
\providecommand{\natexlab}[1]{#1}
{\catcode`\|=0\catcode`\#=12\catcode`\@=11\catcode`\\=12
|immediate|write|@auxout{\expandafter\ifx\csname
  natexlab\endcsname\relax\gdef\natexlab#1{#1}\fi}}
\renewcommand{\url}[1]{{\tt #1}}
\providecommand{\urlprefix}{URL }
\expandafter\ifx\csname urlstyle\endcsname\relax
  \providecommand{\doi}[1]{doi:\discretionary{}{}{}#1}\else
  \providecommand{\doi}{doi:\discretionary{}{}{}\begingroup
  \urlstyle{rm}\Url}\fi
\providecommand{\bibinfo}[2]{#2}
\providecommand{\eprint}[2][]{\url{#2}}

\bibtype{Article}%
\bibitem[Adler(1969)]{Adler:1969gk}
\bibinfo{author}{Adler SL} (\bibinfo{year}{1969}).
\bibinfo{title}{{Axial vector vertex in spinor electrodynamics}}.
\bibinfo{journal}{{\em Phys. Rev.}} \bibinfo{volume}{177}:
  \bibinfo{pages}{2426--2438}. \bibinfo{doi}{\doi{10.1103/PhysRev.177.2426}}.

\bibtype{Article}%
\bibitem[Adler and Bardeen(1969)]{Adler:1969er}
\bibinfo{author}{Adler SL} and  \bibinfo{author}{Bardeen WA}
  (\bibinfo{year}{1969}).
\bibinfo{title}{{Absence of higher order corrections in the anomalous axial
  vector divergence equation}}.
\bibinfo{journal}{{\em Phys. Rev.}} \bibinfo{volume}{182}:
  \bibinfo{pages}{1517--1536}. \bibinfo{doi}{\doi{10.1103/PhysRev.182.1517}}.

\bibtype{Article}%
\bibitem[Beane et al.(2011)]{Beane:2010em}
\bibinfo{author}{Beane SR}, \bibinfo{author}{Detmold W},
  \bibinfo{author}{Orginos K} and  \bibinfo{author}{Savage MJ}
  (\bibinfo{year}{2011}).
\bibinfo{title}{{Nuclear Physics from Lattice QCD}}.
\bibinfo{journal}{{\em Prog. Part. Nucl. Phys.}} \bibinfo{volume}{66}:
  \bibinfo{pages}{1--40}. \bibinfo{doi}{\doi{10.1016/j.ppnp.2010.08.002}}.
\eprint{1004.2935}.

\bibtype{Article}%
\bibitem[Becher and Leutwyler(1999)]{Becher:1999he}
\bibinfo{author}{Becher T} and  \bibinfo{author}{Leutwyler H}
  (\bibinfo{year}{1999}).
\bibinfo{title}{{Baryon chiral perturbation theory in manifestly Lorentz
  invariant form}}.
\bibinfo{journal}{{\em Eur. Phys. J.}} \bibinfo{volume}{9}
  (\bibinfo{number}{4}): \bibinfo{pages}{643--671}.
  \bibinfo{doi}{\doi{10.1007/PL00021673}}.
\eprint{hep-ph/9901384}.

\bibtype{Article}%
\bibitem[Bell and Jackiw(1969)]{Bell:1969ts}
\bibinfo{author}{Bell JS} and  \bibinfo{author}{Jackiw R}
  (\bibinfo{year}{1969}).
\bibinfo{title}{{A PCAC puzzle: $\pi^0 \to \gamma \gamma$ in the $\sigma$
  model}}.
\bibinfo{journal}{{\em Nuovo Cim. A}} \bibinfo{volume}{60}:
  \bibinfo{pages}{47--61}. \bibinfo{doi}{\doi{10.1007/BF02823296}}.

\bibtype{Article}%
\bibitem[Bernard and Mei{\ss}ner(2007)]{Bernard:2006gx}
\bibinfo{author}{Bernard V} and  \bibinfo{author}{Mei{\ss}ner UG}
  (\bibinfo{year}{2007}).
\bibinfo{title}{{Chiral perturbation theory}}.
\bibinfo{journal}{{\em Ann. Rev. Nucl. Part. Sci.}} \bibinfo{volume}{57}:
  \bibinfo{pages}{33--60}.
  \bibinfo{doi}{\doi{10.1146/annurev.nucl.56.080805.140449}}.
\eprint{hep-ph/0611231}.

\bibtype{Article}%
\bibitem[Bernard et al.(1992)]{Bernard:1992qa}
\bibinfo{author}{Bernard V}, \bibinfo{author}{Kaiser N},
  \bibinfo{author}{Kambor J} and  \bibinfo{author}{Mei{\ss}ner UG}
  (\bibinfo{year}{1992}).
\bibinfo{title}{{Chiral structure of the nucleon}}.
\bibinfo{journal}{{\em Nucl. Phys. B}} \bibinfo{volume}{388}:
  \bibinfo{pages}{315--345}. \bibinfo{doi}{\doi{10.1016/0550-3213(92)90615-I}}.

\bibtype{Article}%
\bibitem[Bernard et al.(1995)]{Bernard:1995dp}
\bibinfo{author}{Bernard V}, \bibinfo{author}{Kaiser N} and
  \bibinfo{author}{Mei{\ss}ner UG} (\bibinfo{year}{1995}).
\bibinfo{title}{{Chiral dynamics in nucleons and nuclei}}.
\bibinfo{journal}{{\em Int. J. Mod. Phys. E}} \bibinfo{volume}{4}:
  \bibinfo{pages}{193--346}. \bibinfo{doi}{\doi{10.1142/S0218301395000092}}.
\eprint{hep-ph/9501384}.

\bibtype{Article}%
\bibitem[Bonanno et al.(2025)]{Bonanno:2025wcv}
\bibinfo{author}{Bonanno C}, \bibinfo{author}{Bonati C} and
  \bibinfo{author}{D'Elia M} (\bibinfo{year}{2025}), \bibinfo{month}{10}.
\bibinfo{title}{{Strong CP problem, theta term and QCD topological properties}}
  \eprint{2510.03059}.

\bibtype{Book}%
\bibitem[Brauner(2024)]{Brauner:2024juy}
\bibinfo{author}{Brauner T} (\bibinfo{year}{2024}).
\bibinfo{title}{{Effective Field Theory for Spontaneously Broken Symmetry}},
  \bibinfo{series}{Lect. Notes Phys.}, \bibinfo{volume}{1023},
  \bibinfo{publisher}{Springer}.
\bibinfo{doi}{\doi{10.1007/978-3-031-48378-3}}.
\eprint{2404.14518}.

\bibtype{Article}%
\bibitem[Broussard et al.(2025)]{Broussard:2025opd}
\bibinfo{author}{Broussard LJ} and  et al. (\bibinfo{year}{2025}).
\bibinfo{title}{{Baryon number violation: from nuclear matrix elements to BSM
  physics}}.
\bibinfo{journal}{{\em J. Phys. G}} \bibinfo{volume}{52} (\bibinfo{number}{8}):
  \bibinfo{pages}{083001}. \bibinfo{doi}{\doi{10.1088/1361-6471/adf081}}.
\eprint{2504.16983}.

\bibtype{Book}%
\bibitem[Burgess(2020)]{Burgess:2020tbq}
\bibinfo{author}{Burgess CP} (\bibinfo{year}{2020}), \bibinfo{month}{12}.
\bibinfo{title}{{Introduction to Effective Field Theory}},
  \bibinfo{publisher}{Cambridge University Press}.
\bibinfo{comment}{ISBN} \bibinfo{isbn}{978-1-139-04804-0, 978-0-521-19547-8}.
\bibinfo{doi}{\doi{10.1017/9781139048040}}.

\bibtype{Article}%
\bibitem[Callan et al.(1969)]{Callan:1969sn}
\bibinfo{author}{Callan Jr. CG}, \bibinfo{author}{Coleman SR},
  \bibinfo{author}{Wess J} and  \bibinfo{author}{Zumino B}
  (\bibinfo{year}{1969}).
\bibinfo{title}{{Structure of phenomenological Lagrangians. 2.}}
\bibinfo{journal}{{\em Phys. Rev.}} \bibinfo{volume}{177}:
  \bibinfo{pages}{2247--2250}. \bibinfo{doi}{\doi{10.1103/PhysRev.177.2247}}.

\bibtype{Article}%
\bibitem[Cata and Mateu(2007)]{Cata:2007ns}
\bibinfo{author}{Cata O} and  \bibinfo{author}{Mateu V} (\bibinfo{year}{2007}).
\bibinfo{title}{{Chiral perturbation theory with tensor sources}}.
\bibinfo{journal}{{\em JHEP}} \bibinfo{volume}{09}: \bibinfo{pages}{078}.
  \bibinfo{doi}{\doi{10.1088/1126-6708/2007/09/078}}.
\eprint{0705.2948}.

\bibtype{Article}%
\bibitem[Coleman(1966)]{10.1063/1.1931207}
\bibinfo{author}{Coleman S} (\bibinfo{year}{1966}), \bibinfo{month}{05}.
\bibinfo{title}{The invariance of the vacuum is the invariance of the world}.
\bibinfo{journal}{{\em Journal of Mathematical Physics}} \bibinfo{volume}{7}
  (\bibinfo{number}{5}): \bibinfo{pages}{787--787}.

\bibtype{Book}%
\bibitem[Coleman(1985)]{Coleman:1985rnk}
\bibinfo{author}{Coleman S} (\bibinfo{year}{1985}).
\bibinfo{title}{{Aspects of Symmetry}: {Selected Erice Lectures}},
  \bibinfo{publisher}{Cambridge University Press}, \bibinfo{address}{Cambridge,
  U.K.}
\bibinfo{comment}{ISBN} \bibinfo{isbn}{978-0-521-31827-3}.
\bibinfo{doi}{\doi{10.1017/CBO9780511565045}}.

\bibtype{Article}%
\bibitem[Coleman and Witten(1980)]{Coleman:1980mx}
\bibinfo{author}{Coleman SR} and  \bibinfo{author}{Witten E}
  (\bibinfo{year}{1980}).
\bibinfo{title}{{Chiral Symmetry Breakdown in Large N Chromodynamics}}.
\bibinfo{journal}{{\em Phys. Rev. Lett.}} \bibinfo{volume}{45}:
  \bibinfo{pages}{100}. \bibinfo{doi}{\doi{10.1103/PhysRevLett.45.100}}.

\bibtype{Article}%
\bibitem[Coleman et al.(1969)]{Coleman:1969sm}
\bibinfo{author}{Coleman SR}, \bibinfo{author}{Wess J} and
  \bibinfo{author}{Zumino B} (\bibinfo{year}{1969}).
\bibinfo{title}{{Structure of phenomenological Lagrangians. 1.}}
\bibinfo{journal}{{\em Phys. Rev.}} \bibinfo{volume}{177}:
  \bibinfo{pages}{2239--2247}. \bibinfo{doi}{\doi{10.1103/PhysRev.177.2239}}.

\bibtype{Article}%
\bibitem[Dashen and Manohar(1993{\natexlab{a}})]{Dashen:1993ac}
\bibinfo{author}{Dashen RF} and  \bibinfo{author}{Manohar AV}
  (\bibinfo{year}{1993}{\natexlab{a}}).
\bibinfo{title}{{1/N(c) corrections to the baryon axial currents in QCD}}.
\bibinfo{journal}{{\em Phys. Lett. B}} \bibinfo{volume}{315}:
  \bibinfo{pages}{438--440}. \bibinfo{doi}{\doi{10.1016/0370-2693(93)91637-3}}.
\eprint{hep-ph/9307242}.

\bibtype{Article}%
\bibitem[Dashen and Manohar(1993{\natexlab{b}})]{Dashen:1993as}
\bibinfo{author}{Dashen RF} and  \bibinfo{author}{Manohar AV}
  (\bibinfo{year}{1993}{\natexlab{b}}).
\bibinfo{title}{{Baryon-pion couplings from large N(c) QCD}}.
\bibinfo{journal}{{\em Phys. Lett. B}} \bibinfo{volume}{315}:
  \bibinfo{pages}{425--430}. \bibinfo{doi}{\doi{10.1016/0370-2693(93)91635-Z}}.
\eprint{hep-ph/9307241}.

\bibtype{Article}%
\bibitem[Dashen et al.(1994)]{Dashen:1993jt}
\bibinfo{author}{Dashen RF}, \bibinfo{author}{Jenkins EE} and
  \bibinfo{author}{Manohar AV} (\bibinfo{year}{1994}).
\bibinfo{title}{{The 1/N(c) expansion for baryons}}.
\bibinfo{journal}{{\em Phys. Rev. D}} \bibinfo{volume}{49}:
  \bibinfo{pages}{4713}. \bibinfo{doi}{\doi{10.1103/PhysRevD.51.2489}}.
\bibinfo{note}{[Erratum: Phys.Rev.D 51, 2489 (1995)]}, \eprint{hep-ph/9310379}.

\bibtype{Book}%
\bibitem[Donoghue et al.(2022)]{Donoghue:2022wrw}
\bibinfo{author}{Donoghue JF}, \bibinfo{author}{Golowich E} and
  \bibinfo{author}{Holstein BR} (\bibinfo{year}{2022}), \bibinfo{month}{11}.
\bibinfo{title}{{Dynamics of the Standard Model}: {Second edition}},
  \bibinfo{publisher}{Cambridge University Press}.
\bibinfo{comment}{ISBN} \bibinfo{isbn}{978-1-009-29100-2, 978-1-009-29101-9,
  978-1-009-29103-3}.
\bibinfo{doi}{\doi{10.1017/9781009291033}}.

\bibtype{Article}%
\bibitem[Drischler et al.(2021)]{Drischler:2019xuo}
\bibinfo{author}{Drischler C}, \bibinfo{author}{Haxton W},
  \bibinfo{author}{McElvain K}, \bibinfo{author}{Mereghetti E},
  \bibinfo{author}{Nicholson A}, \bibinfo{author}{Vranas P} and
  \bibinfo{author}{Walker-Loud A} (\bibinfo{year}{2021}).
\bibinfo{title}{{Towards grounding nuclear physics in QCD}}.
\bibinfo{journal}{{\em Prog. Part. Nucl. Phys.}} \bibinfo{volume}{121}:
  \bibinfo{pages}{103888}. \bibinfo{doi}{\doi{10.1016/j.ppnp.2021.103888}}.
\eprint{1910.07961}.

\bibtype{Article}%
\bibitem[Eichten and Hill(1990)]{Eichten:1989zv}
\bibinfo{author}{Eichten E} and  \bibinfo{author}{Hill BR}
  (\bibinfo{year}{1990}).
\bibinfo{title}{{An Effective Field Theory for the Calculation of Matrix
  Elements Involving Heavy Quarks}}.
\bibinfo{journal}{{\em Phys. Lett. B}} \bibinfo{volume}{234}:
  \bibinfo{pages}{511--516}. \bibinfo{doi}{\doi{10.1016/0370-2693(90)92049-O}}.

\bibtype{Misc}%
\bibitem[Enc(2026)]{EncycPartHQET}
 (\bibinfo{year}{2026}).
\bibinfo{title}{{Effective field theories for heavy quarks}}.
\bibinfo{howpublished}{Encyclopedia of Particle Physics}.

\bibtype{Article}%
\bibitem[Englert and Brout(1964)]{Englert:1964et}
\bibinfo{author}{Englert F} and  \bibinfo{author}{Brout R}
  (\bibinfo{year}{1964}).
\bibinfo{title}{{Broken Symmetry and the Mass of Gauge Vector Mesons}}.
\bibinfo{journal}{{\em Phys. Rev. Lett.}} \bibinfo{volume}{13}:
  \bibinfo{pages}{321--323}. \bibinfo{doi}{\doi{10.1103/PhysRevLett.13.321}}.

\bibtype{Article}%
\bibitem[Epelbaum et al.(2015)]{Epelbaum:2014sea}
\bibinfo{author}{Epelbaum E}, \bibinfo{author}{Gasparyan AM},
  \bibinfo{author}{Krebs H} and  \bibinfo{author}{Schat C}
  (\bibinfo{year}{2015}).
\bibinfo{title}{{Three-nucleon force at large distances: Insights from chiral
  effective field theory and the large-N$_{c}$ expansion}}.
\bibinfo{journal}{{\em Eur. Phys. J. A}} \bibinfo{volume}{51}
  (\bibinfo{number}{3}): \bibinfo{pages}{26}.
  \bibinfo{doi}{\doi{10.1140/epja/i2015-15026-y}}.
\eprint{1411.3612}.

\bibtype{Article}%
\bibitem[Faber and H{\"o}llwieser(2017)]{Faber:2017alm}
\bibinfo{author}{Faber M} and  \bibinfo{author}{H{\"o}llwieser R}
  (\bibinfo{year}{2017}).
\bibinfo{title}{{Chiral symmetry breaking on the lattice}}.
\bibinfo{journal}{{\em Prog. Part. Nucl. Phys.}} \bibinfo{volume}{97}:
  \bibinfo{pages}{312--355}. \bibinfo{doi}{\doi{10.1016/j.ppnp.2017.08.001}}.
\eprint{1908.09740}.

\bibtype{Article}%
\bibitem[Fettes et al.(2000)]{Fettes:2000gb}
\bibinfo{author}{Fettes N}, \bibinfo{author}{Mei{\ss}ner UG},
  \bibinfo{author}{Mojzis M} and  \bibinfo{author}{Steininger S}
  (\bibinfo{year}{2000}).
\bibinfo{title}{{The chiral effective pion nucleon Lagrangian of order p**4}}.
\bibinfo{journal}{{\em Annals Phys.}} \bibinfo{volume}{283}:
  \bibinfo{pages}{273--302}. \bibinfo{doi}{\doi{10.1006/aphy.2000.6059}}.
\bibinfo{note}{[Erratum: Annals Phys. 288, 249--250 (2001)]},
  \eprint{hep-ph/0001308}.

\bibtype{Article}%
\bibitem[Fileviez~Perez et al.(2022)]{FileviezPerez:2022ypk}
\bibinfo{author}{Fileviez~Perez P} and  et al. (\bibinfo{year}{2022}),
  \bibinfo{month}{7}.
\bibinfo{title}{{On Baryon and Lepton Number Violation}} \eprint{2208.00010}.

\bibtype{Article}%
\bibitem[Foldy and Wouthuysen(1950)]{Foldy:1949wa}
\bibinfo{author}{Foldy LL} and  \bibinfo{author}{Wouthuysen SA}
  (\bibinfo{year}{1950}).
\bibinfo{title}{{On the Dirac theory of spin 1/2 particle and its
  nonrelativistic limit}}.
\bibinfo{journal}{{\em Phys. Rev.}} \bibinfo{volume}{78}:
  \bibinfo{pages}{29--36}. \bibinfo{doi}{\doi{10.1103/PhysRev.78.29}}.

\bibtype{Article}%
\bibitem[Fritzsch et al.(1973)]{Fritzsch:1973pi}
\bibinfo{author}{Fritzsch H}, \bibinfo{author}{Gell-Mann M} and
  \bibinfo{author}{Leutwyler H} (\bibinfo{year}{1973}).
\bibinfo{title}{{Advantages of the Color Octet Gluon Picture}}.
\bibinfo{journal}{{\em Phys. Lett. B}} \bibinfo{volume}{47}:
  \bibinfo{pages}{365--368}. \bibinfo{doi}{\doi{10.1016/0370-2693(73)90625-4}}.

\bibtype{Article}%
\bibitem[Fuchs et al.(2003)]{Fuchs:2003qc}
\bibinfo{author}{Fuchs T}, \bibinfo{author}{Gegelia J},
  \bibinfo{author}{Japaridze G} and  \bibinfo{author}{Scherer S}
  (\bibinfo{year}{2003}).
\bibinfo{title}{{Renormalization of relativistic baryon chiral perturbation
  theory and power counting}}.
\bibinfo{journal}{{\em Phys. Rev. D}} \bibinfo{volume}{68}:
  \bibinfo{pages}{056005}. \bibinfo{doi}{\doi{10.1103/PhysRevD.68.056005}}.
\eprint{hep-ph/0302117}.

\bibtype{Article}%
\bibitem[Fuchs et al.(2004)]{Fuchs:2003kq}
\bibinfo{author}{Fuchs T}, \bibinfo{author}{Gegelia J} and
  \bibinfo{author}{Scherer S} (\bibinfo{year}{2004}).
\bibinfo{title}{{Structure of the nucleon in chiral perturbation theory}}.
\bibinfo{journal}{{\em Eur. Phys. J. A}} \bibinfo{volume}{19}:
  \bibinfo{pages}{35--42}. \bibinfo{doi}{\doi{10.1140/epjad/s2004-03-006-0}}.
\eprint{hep-ph/0309234}.

\bibtype{Misc}%
\bibitem[Gardner et al.(2026)]{EncycPV}
\bibinfo{author}{Gardner S} and  et al. (\bibinfo{year}{2026}).
\bibinfo{title}{{Hadronic parity violation}}.
\bibinfo{howpublished}{This volume}.

\bibtype{Article}%
\bibitem[Gasser and Leutwyler(1984)]{Gasser:1983yg}
\bibinfo{author}{Gasser J} and  \bibinfo{author}{Leutwyler H}
  (\bibinfo{year}{1984}).
\bibinfo{title}{{Chiral Perturbation Theory to One Loop}}.
\bibinfo{journal}{{\em Annals Phys.}} \bibinfo{volume}{158}:
  \bibinfo{pages}{142}. \bibinfo{doi}{\doi{10.1016/0003-4916(84)90242-2}}.

\bibtype{Article}%
\bibitem[Gasser and Leutwyler(1985)]{Gasser:1984gg}
\bibinfo{author}{Gasser J} and  \bibinfo{author}{Leutwyler H}
  (\bibinfo{year}{1985}).
\bibinfo{title}{{Chiral Perturbation Theory: Expansions in the Mass of the
  Strange Quark}}.
\bibinfo{journal}{{\em Nucl. Phys. B}} \bibinfo{volume}{250}:
  \bibinfo{pages}{465--516}. \bibinfo{doi}{\doi{10.1016/0550-3213(85)90492-4}}.

\bibtype{Article}%
\bibitem[Gasser et al.(1988)]{Gasser:1987rb}
\bibinfo{author}{Gasser J}, \bibinfo{author}{Sainio ME} and
  \bibinfo{author}{Svarc A} (\bibinfo{year}{1988}).
\bibinfo{title}{{Nucleons with chiral loops}}.
\bibinfo{journal}{{\em Nucl. Phys. B}} \bibinfo{volume}{307}:
  \bibinfo{pages}{779--853}. \bibinfo{doi}{\doi{10.1016/0550-3213(88)90108-3}}.

\bibtype{Article}%
\bibitem[Gegelia and Japaridze(1999)]{Gegelia:1999gf}
\bibinfo{author}{Gegelia J} and  \bibinfo{author}{Japaridze G}
  (\bibinfo{year}{1999}).
\bibinfo{title}{{Matching heavy particle approach to relativistic theory}}.
\bibinfo{journal}{{\em Phys. Rev. D}} \bibinfo{volume}{60}:
  \bibinfo{pages}{114038}. \bibinfo{doi}{\doi{10.1103/PhysRevD.60.114038}}.
\eprint{hep-ph/9908377}.

\bibtype{Techreport}%
\bibitem[Gell-Mann(1961)]{Gell-Mann:1961omu}
\bibinfo{author}{Gell-Mann M}.
\bibinfo{title}{The eightfold way: A theory of strong interaction symmetry}. ,
  \bibinfo{institution}{California Inst. of Tech., Pasadena. Synchrotron Lab.}
  (\bibinfo{year}{1961}).
\bibinfo{doi}{\doi{10.2172/4008239}}.
  \bibinfo{url}{\url{https://www.osti.gov/biblio/4008239}}.

\bibtype{Article}%
\bibitem[Gell-Mann(1962)]{Gell-Mann:1962yej}
\bibinfo{author}{Gell-Mann M} (\bibinfo{year}{1962}).
\bibinfo{title}{{Symmetries of baryons and mesons}}.
\bibinfo{journal}{{\em Phys. Rev.}} \bibinfo{volume}{125}:
  \bibinfo{pages}{1067--1084}. \bibinfo{doi}{\doi{10.1103/PhysRev.125.1067}}.

\bibtype{Article}%
\bibitem[Gell-Mann et al.(1968)]{Gell-Mann:1968hlm}
\bibinfo{author}{Gell-Mann M}, \bibinfo{author}{Oakes RJ} and
  \bibinfo{author}{Renner B} (\bibinfo{year}{1968}).
\bibinfo{title}{{Behavior of current divergences under SU(3) x SU(3)}}.
\bibinfo{journal}{{\em Phys. Rev.}} \bibinfo{volume}{175}:
  \bibinfo{pages}{2195--2199}. \bibinfo{doi}{\doi{10.1103/PhysRev.175.2195}}.

\bibtype{Article}%
\bibitem[Georgi(1990)]{Georgi:1990um}
\bibinfo{author}{Georgi H} (\bibinfo{year}{1990}).
\bibinfo{title}{{An Effective Field Theory for Heavy Quarks at Low Energies}}.
\bibinfo{journal}{{\em Phys. Lett. B}} \bibinfo{volume}{240}:
  \bibinfo{pages}{447--450}. \bibinfo{doi}{\doi{10.1016/0370-2693(90)91128-X}}.

\bibtype{Book}%
\bibitem[Georgi(2009)]{Georgi:WeakInt}
\bibinfo{author}{Georgi H} (\bibinfo{year}{2009}).
\bibinfo{title}{{Weak Interactions and Modern Particle Theory}},
  \bibinfo{publisher}{Dover Publications}.
\bibinfo{comment}{ISBN} \bibinfo{isbn}{9780486469041}.

\bibtype{Article}%
\bibitem[Gervais and Sakita(1984{\natexlab{a}})]{Gervais:1984rc}
\bibinfo{author}{Gervais JL} and  \bibinfo{author}{Sakita B}
  (\bibinfo{year}{1984}{\natexlab{a}}).
\bibinfo{title}{{Large N Baryonic Soliton and Quarks}}.
\bibinfo{journal}{{\em Phys. Rev. D}} \bibinfo{volume}{30}:
  \bibinfo{pages}{1795}. \bibinfo{doi}{\doi{10.1103/PhysRevD.30.1795}}.

\bibtype{Article}%
\bibitem[Gervais and Sakita(1984{\natexlab{b}})]{Gervais:1983wq}
\bibinfo{author}{Gervais JL} and  \bibinfo{author}{Sakita B}
  (\bibinfo{year}{1984}{\natexlab{b}}).
\bibinfo{title}{{Large N QCD Baryon Dynamics: Exact Results from Its Relation
  to the Static Strong Coupling Theory}}.
\bibinfo{journal}{{\em Phys. Rev. Lett.}} \bibinfo{volume}{52}:
  \bibinfo{pages}{87}. \bibinfo{doi}{\doi{10.1103/PhysRevLett.52.87}}.

\bibtype{Article}%
\bibitem[Goldstone(1961)]{Goldstone:1961eq}
\bibinfo{author}{Goldstone J} (\bibinfo{year}{1961}).
\bibinfo{title}{{Field Theories with Superconductor Solutions}}.
\bibinfo{journal}{{\em Nuovo Cim.}} \bibinfo{volume}{19}:
  \bibinfo{pages}{154--164}. \bibinfo{doi}{\doi{10.1007/BF02812722}}.

\bibtype{Article}%
\bibitem[Gross and Wilczek(1973)]{Gross:1973id}
\bibinfo{author}{Gross DJ} and  \bibinfo{author}{Wilczek F}
  (\bibinfo{year}{1973}).
\bibinfo{title}{{Ultraviolet Behavior of Nonabelian Gauge Theories}}.
\bibinfo{journal}{{\em Phys. Rev. Lett.}} \bibinfo{volume}{30}:
  \bibinfo{pages}{1343--1346}.
  \bibinfo{doi}{\doi{10.1103/PhysRevLett.30.1343}}.

\bibtype{Article}%
\bibitem[Gross et al.(2023)]{Gross:2022hyw}
\bibinfo{author}{Gross F} and  et al. (\bibinfo{year}{2023}).
\bibinfo{title}{{50 Years of Quantum Chromodynamics}}.
\bibinfo{journal}{{\em Eur. Phys. J. C}} \bibinfo{volume}{83}:
  \bibinfo{pages}{1125}. \bibinfo{doi}{\doi{10.1140/epjc/s10052-023-11949-2}}.
\eprint{2212.11107}.

\bibtype{Article}%
\bibitem[Guralnik et al.(1964)]{Guralnik:1964eu}
\bibinfo{author}{Guralnik GS}, \bibinfo{author}{Hagen CR} and
  \bibinfo{author}{Kibble TWB} (\bibinfo{year}{1964}).
\bibinfo{title}{{Global Conservation Laws and Massless Particles}}.
\bibinfo{journal}{{\em Phys. Rev. Lett.}} \bibinfo{volume}{13}:
  \bibinfo{pages}{585--587}. \bibinfo{doi}{\doi{10.1103/PhysRevLett.13.585}}.

\bibtype{Article}%
\bibitem[Haidenbauer and Mei{\ss}ner(2022)]{Haidenbauer:2022esw}
\bibinfo{author}{Haidenbauer J} and  \bibinfo{author}{Mei{\ss}ner UG}
  (\bibinfo{year}{2022}).
\bibinfo{title}{{Status of the hyperon-nucleon interaction in chiral effective
  field theory}}.
\bibinfo{journal}{{\em EPJ Web Conf.}} \bibinfo{volume}{271}:
  \bibinfo{pages}{05001}. \bibinfo{doi}{\doi{10.1051/epjconf/202227105001}}.
\eprint{2208.13542}.

\bibtype{Misc}%
\bibitem[Hammer(2026)]{EncycEFT}
\bibinfo{author}{Hammer HW} (\bibinfo{year}{2026}).
\bibinfo{title}{{Effective field theory and renormalization}}.
\bibinfo{howpublished}{This volume}.

\bibtype{Article}%
\bibitem[Hammer et al.(2020)]{Hammer:2019poc}
\bibinfo{author}{Hammer HW}, \bibinfo{author}{K{\"o}nig S} and
  \bibinfo{author}{van Kolck U} (\bibinfo{year}{2020}).
\bibinfo{title}{{Nuclear effective field theory: status and perspectives}}.
\bibinfo{journal}{{\em Rev. Mod. Phys.}} \bibinfo{volume}{92}
  (\bibinfo{number}{2}): \bibinfo{pages}{025004}.
  \bibinfo{doi}{\doi{10.1103/RevModPhys.92.025004}}.
\eprint{1906.12122}.

\bibtype{Article}%
\bibitem[Haxton and Holstein(2013)]{Haxton:2013aca}
\bibinfo{author}{Haxton WC} and  \bibinfo{author}{Holstein BR}
  (\bibinfo{year}{2013}).
\bibinfo{title}{{Hadronic Parity Violation}}.
\bibinfo{journal}{{\em Prog. Part. Nucl. Phys.}} \bibinfo{volume}{71}:
  \bibinfo{pages}{185--203}. \bibinfo{doi}{\doi{10.1016/j.ppnp.2013.03.009}}.
\eprint{1303.4132}.

\bibtype{Misc}%
\bibitem[Hebeler et al.(2026)]{EncycNN}
\bibinfo{author}{Hebeler K} and  et al. (\bibinfo{year}{2026}).
\bibinfo{title}{{Chiral EFT for nuclear forces}}.
\bibinfo{howpublished}{This volume}.

\bibtype{Article}%
\bibitem[Heisenberg(1932)]{Heisenberg:iso}
\bibinfo{author}{Heisenberg W} (\bibinfo{year}{1932}).
\bibinfo{title}{{{\"U}ber den Bau der Atomkerne. I}}.
\bibinfo{journal}{{\em Zeitschrift f{\"u}r Physik}} \bibinfo{volume}{77}
  (\bibinfo{number}{1}): \bibinfo{pages}{1--11}.

\bibtype{Article}%
\bibitem[Higgs(1964)]{Higgs:1964pj}
\bibinfo{author}{Higgs PW} (\bibinfo{year}{1964}).
\bibinfo{title}{{Broken Symmetries and the Masses of Gauge Bosons}}.
\bibinfo{journal}{{\em Phys. Rev. Lett.}} \bibinfo{volume}{13}:
  \bibinfo{pages}{508--509}. \bibinfo{doi}{\doi{10.1103/PhysRevLett.13.508}}.

\bibtype{Article}%
\bibitem[Hoferichter et al.(2016)]{Hoferichter:2015hva}
\bibinfo{author}{Hoferichter M}, \bibinfo{author}{Ruiz~de Elvira J},
  \bibinfo{author}{Kubis B} and  \bibinfo{author}{Mei{\ss}ner UG}
  (\bibinfo{year}{2016}).
\bibinfo{title}{{Roy{\textendash}Steiner-equation analysis of
  pion{\textendash}nucleon scattering}}.
\bibinfo{journal}{{\em Phys. Rept.}} \bibinfo{volume}{625}:
  \bibinfo{pages}{1--88}. \bibinfo{doi}{\doi{10.1016/j.physrep.2016.02.002}}.
\eprint{1510.06039}.

\bibtype{Article}%
\bibitem[Hook(2019)]{Hook:2018dlk}
\bibinfo{author}{Hook A} (\bibinfo{year}{2019}).
\bibinfo{title}{{TASI Lectures on the Strong CP Problem and Axions}}.
\bibinfo{journal}{{\em PoS}} \bibinfo{volume}{TASI2018}: \bibinfo{pages}{004}.
  \bibinfo{doi}{\doi{10.22323/1.333.0004}}.
\eprint{1812.02669}.

\bibtype{Article}%
\bibitem[Isgur and Wise(1989)]{Isgur:1989vq}
\bibinfo{author}{Isgur N} and  \bibinfo{author}{Wise MB}
  (\bibinfo{year}{1989}).
\bibinfo{title}{{Weak Decays of Heavy Mesons in the Static Quark
  Approximation}}.
\bibinfo{journal}{{\em Phys. Lett. B}} \bibinfo{volume}{232}:
  \bibinfo{pages}{113--117}. \bibinfo{doi}{\doi{10.1016/0370-2693(89)90566-2}}.

\bibtype{Article}%
\bibitem[Jenkins(1998)]{Jenkins:1998wy}
\bibinfo{author}{Jenkins EE} (\bibinfo{year}{1998}).
\bibinfo{title}{{Large N(c) baryons}}.
\bibinfo{journal}{{\em Ann. Rev. Nucl. Part. Sci.}} \bibinfo{volume}{48}:
  \bibinfo{pages}{81--119}. \bibinfo{doi}{\doi{10.1146/annurev.nucl.48.1.81}}.
\eprint{hep-ph/9803349}.

\bibtype{Article}%
\bibitem[Jenkins and Manohar(1991)]{Jenkins:1990jv}
\bibinfo{author}{Jenkins EE} and  \bibinfo{author}{Manohar AV}
  (\bibinfo{year}{1991}).
\bibinfo{title}{{Baryon chiral perturbation theory using a heavy fermion
  Lagrangian}}.
\bibinfo{journal}{{\em Phys. Lett. B}} \bibinfo{volume}{255}:
  \bibinfo{pages}{558--562}. \bibinfo{doi}{\doi{10.1016/0370-2693(91)90266-S}}.

\bibtype{Article}%
\bibitem[Kaiser and Leutwyler(2000)]{Kaiser:2000gs}
\bibinfo{author}{Kaiser R} and  \bibinfo{author}{Leutwyler H}
  (\bibinfo{year}{2000}).
\bibinfo{title}{{Large N(c) in chiral perturbation theory}}.
\bibinfo{journal}{{\em Eur. Phys. J. C}} \bibinfo{volume}{17}:
  \bibinfo{pages}{623--649}. \bibinfo{doi}{\doi{10.1007/s100520000499}}.
\eprint{hep-ph/0007101}.

\bibtype{Article}%
\bibitem[Kaplan and Manohar(1997)]{Kaplan:1996rk}
\bibinfo{author}{Kaplan DB} and  \bibinfo{author}{Manohar AV}
  (\bibinfo{year}{1997}).
\bibinfo{title}{{The nucleon-nucleon potential in the 1/N(c) expansion}}.
\bibinfo{journal}{{\em Phys. Rev. C}} \bibinfo{volume}{56}:
  \bibinfo{pages}{76--83}. \bibinfo{doi}{\doi{10.1103/PhysRevC.56.76}}.
\eprint{nucl-th/9612021}.

\bibtype{Article}%
\bibitem[Kaplan and Savage(1996)]{Kaplan:1995yg}
\bibinfo{author}{Kaplan DB} and  \bibinfo{author}{Savage MJ}
  (\bibinfo{year}{1996}).
\bibinfo{title}{{The spin flavor dependence of nuclear forces from large-N
  QCD}}.
\bibinfo{journal}{{\em Phys. Lett. B}} \bibinfo{volume}{365}:
  \bibinfo{pages}{244--251}. \bibinfo{doi}{\doi{10.1016/0370-2693(95)01277-X}}.
\eprint{hep-ph/9509371}.

\bibtype{Article}%
\bibitem[Kim and Carosi(2010)]{Kim:2008hd}
\bibinfo{author}{Kim JE} and  \bibinfo{author}{Carosi G}
  (\bibinfo{year}{2010}).
\bibinfo{title}{{Axions and the Strong CP Problem}}.
\bibinfo{journal}{{\em Rev. Mod. Phys.}} \bibinfo{volume}{82}:
  \bibinfo{pages}{557--602}. \bibinfo{doi}{\doi{10.1103/RevModPhys.82.557}}.
\bibinfo{note}{[Erratum: Rev.Mod.Phys. 91, 049902 (2019)]}, \eprint{0807.3125}.

\bibtype{Article}%
\bibitem[Lebed(1999)]{Lebed:1998st}
\bibinfo{author}{Lebed RF} (\bibinfo{year}{1999}).
\bibinfo{title}{{Phenomenology of large N(c) QCD}}.
\bibinfo{journal}{{\em Czech. J. Phys.}} \bibinfo{volume}{49}:
  \bibinfo{pages}{1273--1306}. \bibinfo{doi}{\doi{10.1023/A:1022820227262}}.
\eprint{nucl-th/9810080}.

\bibtype{Article}%
\bibitem[Lee et al.(2021)]{Lee:2020esp}
\bibinfo{author}{Lee D} and  et al. (\bibinfo{year}{2021}).
\bibinfo{title}{{Hidden Spin-Isospin Exchange Symmetry}}.
\bibinfo{journal}{{\em Phys. Rev. Lett.}} \bibinfo{volume}{127}
  (\bibinfo{number}{6}): \bibinfo{pages}{062501}.
  \bibinfo{doi}{\doi{10.1103/PhysRevLett.127.062501}}.
\eprint{2010.09420}.

\bibtype{Article}%
\bibitem[Leibbrandt(1975)]{Leibbrandt:1975dj}
\bibinfo{author}{Leibbrandt G} (\bibinfo{year}{1975}).
\bibinfo{title}{{Introduction to the Technique of Dimensional Regularization}}.
\bibinfo{journal}{{\em Rev. Mod. Phys.}} \bibinfo{volume}{47}:
  \bibinfo{pages}{849}. \bibinfo{doi}{\doi{10.1103/RevModPhys.47.849}}.

\bibtype{Article}%
\bibitem[Leutwyler(1994)]{Leutwyler:1993iq}
\bibinfo{author}{Leutwyler H} (\bibinfo{year}{1994}).
\bibinfo{title}{{On the foundations of chiral perturbation theory}}.
\bibinfo{journal}{{\em Annals Phys.}} \bibinfo{volume}{235}:
  \bibinfo{pages}{165--203}. \bibinfo{doi}{\doi{10.1006/aphy.1994.1094}}.
\eprint{hep-ph/9311274}.

\bibtype{Article}%
\bibitem[Leutwyler(1996)]{Leutwyler:1996sa}
\bibinfo{author}{Leutwyler H} (\bibinfo{year}{1996}).
\bibinfo{title}{{Bounds on the light quark masses}}.
\bibinfo{journal}{{\em Phys. Lett. B}} \bibinfo{volume}{374}:
  \bibinfo{pages}{163--168}. \bibinfo{doi}{\doi{10.1016/0370-2693(96)85876-X}}.
\eprint{hep-ph/9601234}.

\bibtype{Article}%
\bibitem[Li~Muli et al.(2025)]{LiMuli:2025zro}
\bibinfo{author}{Li~Muli SS}, \bibinfo{author}{Dj{\"a}rv TR},
  \bibinfo{author}{Forss{\'e}n C} and  \bibinfo{author}{Phillips DR}
  (\bibinfo{year}{2025}), \bibinfo{month}{3}.
\bibinfo{title}{{The role of spin-isospin symmetries in nuclear
  $\beta$-decays}} \eprint{2503.16372}.

\bibtype{Book}%
\bibitem[Lin and Meyer(2015)]{Lin:2015dga}
\bibinfo{editor}{Lin HW} and  \bibinfo{editor}{Meyer HB}, (Eds.)
  (\bibinfo{year}{2015}).
\bibinfo{title}{{Lattice QCD for Nuclear Physics}}, \bibinfo{series}{Lect.
  Notes Phys.}, \bibinfo{volume}{889}, \bibinfo{publisher}{Springer}.
\bibinfo{comment}{ISBN} \bibinfo{isbn}{978-3-319-08021-5, 978-3-319-08022-2}.
\bibinfo{doi}{\doi{10.1007/978-3-319-08022-2}}.

\bibtype{Article}%
\bibitem[Lin et al.(2023)]{Lin:2022yaf}
\bibinfo{author}{Lin X}, \bibinfo{author}{Singh H}, \bibinfo{author}{Springer
  RP} and  \bibinfo{author}{Vanasse J} (\bibinfo{year}{2023}).
\bibinfo{title}{{Cold neutron-deuteron capture and Wigner-SU(4) symmetry}}.
\bibinfo{journal}{{\em Phys. Rev. C}} \bibinfo{volume}{108}
  (\bibinfo{number}{4}): \bibinfo{pages}{044001}.
  \bibinfo{doi}{\doi{10.1103/PhysRevC.108.044001}}.
\eprint{2210.15650}.

\bibtype{Article}%
\bibitem[Machleidt and Sammarruca(2024)]{Machleidt:2024bwl}
\bibinfo{author}{Machleidt R} and  \bibinfo{author}{Sammarruca F}
  (\bibinfo{year}{2024}).
\bibinfo{title}{{Recent advances in chiral EFT based nuclear forces and their
  applications}}.
\bibinfo{journal}{{\em Prog. Part. Nucl. Phys.}} \bibinfo{volume}{137}:
  \bibinfo{pages}{104117}. \bibinfo{doi}{\doi{10.1016/j.ppnp.2024.104117}}.
\eprint{2402.14032}.

\bibtype{Inproceedings}%
\bibitem[Manohar(1998)]{Manohar:1998xv}
\bibinfo{author}{Manohar AV} (\bibinfo{year}{1998}), \bibinfo{month}{2},
  \bibinfo{title}{{Large N QCD}}, \bibinfo{booktitle}{{Les Houches Summer
  School in Theoretical Physics, Session 68: Probing the Standard Model of
  Particle Interactions}},  \bibinfo{pages}{1091--1169},
  \eprint{hep-ph/9802419}.

\bibtype{Article}%
\bibitem[Mei{\ss}ner(2024)]{Meissner:2024ona}
\bibinfo{author}{Mei{\ss}ner UG} (\bibinfo{year}{2024}), \bibinfo{month}{10}.
\bibinfo{title}{{Chiral perturbation theory}} \eprint{2410.21912}.

\bibtype{Book}%
\bibitem[Mei{\ss}ner and Rusetsky(2022)]{Meissner:2022cbi}
\bibinfo{author}{Mei{\ss}ner UG} and  \bibinfo{author}{Rusetsky A}
  (\bibinfo{year}{2022}), \bibinfo{month}{8}.
\bibinfo{title}{{Effective Field Theories}}, \bibinfo{publisher}{Cambridge
  University Press}.
\bibinfo{comment}{ISBN} \bibinfo{isbn}{978-1-108-68903-8}.
\bibinfo{doi}{\doi{10.1017/9781108689038}}.

\bibtype{Article}%
\bibitem[Moussallam(1995)]{Moussallam:1994xp}
\bibinfo{author}{Moussallam B} (\bibinfo{year}{1995}).
\bibinfo{title}{{Chiral sum rules for parameters of the order six Lagrangian in
  the W-Z sector and application to pi0, eta, eta-prime decays}}.
\bibinfo{journal}{{\em Phys. Rev. D}} \bibinfo{volume}{51}:
  \bibinfo{pages}{4939--4949}. \bibinfo{doi}{\doi{10.1103/PhysRevD.51.4939}}.
\eprint{hep-ph/9407402}.

\bibtype{Article}%
\bibitem[Nambu(1960{\natexlab{a}})]{Nambu:1960xd}
\bibinfo{author}{Nambu Y} (\bibinfo{year}{1960}{\natexlab{a}}).
\bibinfo{title}{{Axial vector current conservation in weak interactions}}.
\bibinfo{journal}{{\em Phys. Rev. Lett.}} \bibinfo{volume}{4}:
  \bibinfo{pages}{380--382}. \bibinfo{doi}{\doi{10.1103/PhysRevLett.4.380}}.

\bibtype{Article}%
\bibitem[Nambu(1960{\natexlab{b}})]{Nambu:1960tm}
\bibinfo{author}{Nambu Y} (\bibinfo{year}{1960}{\natexlab{b}}).
\bibinfo{title}{{Quasiparticles and Gauge Invariance in the Theory of
  Superconductivity}}.
\bibinfo{journal}{{\em Phys. Rev.}} \bibinfo{volume}{117}:
  \bibinfo{pages}{648--663}. \bibinfo{doi}{\doi{10.1103/PhysRev.117.648}}.

\bibtype{Article}%
\bibitem[Nambu and Jona-Lasinio(1961)]{Nambu:1961tp}
\bibinfo{author}{Nambu Y} and  \bibinfo{author}{Jona-Lasinio G}
  (\bibinfo{year}{1961}).
\bibinfo{title}{{Dynamical Model of Elementary Particles Based on an Analogy
  with Superconductivity. 1.}}
\bibinfo{journal}{{\em Phys. Rev.}} \bibinfo{volume}{122}:
  \bibinfo{pages}{345--358}. \bibinfo{doi}{\doi{10.1103/PhysRev.122.345}}.

\bibtype{Article}%
\bibitem[Navas et al.(2024)]{PDG}
\bibinfo{author}{Navas S} and  et al. (\bibinfo{collaboration}{Particle Data
  Group}) (\bibinfo{year}{2024}).
\bibinfo{title}{{Review of particle physics}}.
\bibinfo{journal}{{\em Phys. Rev. D}} \bibinfo{volume}{110}
  (\bibinfo{number}{3}): \bibinfo{pages}{030001}.
  \bibinfo{doi}{\doi{10.1103/PhysRevD.110.030001}}.

\bibtype{Article}%
\bibitem[Ne'eman(1961)]{Neeman:1961jhl}
\bibinfo{author}{Ne'eman Y} (\bibinfo{year}{1961}).
\bibinfo{title}{{Derivation of strong interactions from a gauge invariance}}.
\bibinfo{journal}{{\em Nucl. Phys.}} \bibinfo{volume}{26}:
  \bibinfo{pages}{222--229}. \bibinfo{doi}{\doi{10.1016/0029-5582(61)90134-1}}.

\bibtype{Article}%
\bibitem[Nguyen and Vanasse(2024)]{Nguyen:2024rlr}
\bibinfo{author}{Nguyen HS} and  \bibinfo{author}{Vanasse J}
  (\bibinfo{year}{2024}).
\bibinfo{title}{{Tritium {\ensuremath{\beta}} decay and proton-proton fusion in
  pionless effective field theory}}.
\bibinfo{journal}{{\em Phys. Rev. C}} \bibinfo{volume}{110}
  (\bibinfo{number}{2}): \bibinfo{pages}{L021001}.
  \bibinfo{doi}{\doi{10.1103/PhysRevC.110.L021001}}.
\eprint{2405.07889}.

\bibtype{Misc}%
\bibitem[Nicholson(2026)]{EncycLattice}
\bibinfo{author}{Nicholson A} (\bibinfo{year}{2026}).
\bibinfo{title}{{Nuclear forces from lattice QCD}}.
\bibinfo{howpublished}{This volume}.

\bibtype{Article}%
\bibitem[Noether(1918)]{Noether:1918zz}
\bibinfo{author}{Noether E} (\bibinfo{year}{1918}).
\bibinfo{title}{{Invariant Variation Problems}}.
\bibinfo{journal}{{\em Gott. Nachr.}} \bibinfo{volume}{1918}:
  \bibinfo{pages}{235--257}.

\bibtype{Book}%
\bibitem[Petrov and Blechman(2016)]{Petrov:2016azi}
\bibinfo{author}{Petrov AA} and  \bibinfo{author}{Blechman AE}
  (\bibinfo{year}{2016}).
\bibinfo{title}{{Effective Field Theories}}, \bibinfo{publisher}{World
  Scientific}.
\bibinfo{comment}{ISBN} \bibinfo{isbn}{978-981-4434-92-8, 978-981-4434-94-2}.
\bibinfo{doi}{\doi{10.1142/8619}}.

\bibtype{Article}%
\bibitem[Petschauer et al.(2020)]{Petschauer:2020urh}
\bibinfo{author}{Petschauer S}, \bibinfo{author}{Haidenbauer J},
  \bibinfo{author}{Kaiser N}, \bibinfo{author}{Mei{\ss}ner UG} and
  \bibinfo{author}{Weise W} (\bibinfo{year}{2020}).
\bibinfo{title}{{Hyperon-nuclear interactions from SU(3) chiral effective field
  theory}}.
\bibinfo{journal}{{\em Front. in Phys.}} \bibinfo{volume}{8}:
  \bibinfo{pages}{12}. \bibinfo{doi}{\doi{10.3389/fphy.2020.00012}}.
\eprint{2002.00424}.

\bibtype{Article}%
\bibitem[Phillips and Schat(2013)]{Phillips:2013rsa}
\bibinfo{author}{Phillips DR} and  \bibinfo{author}{Schat C}
  (\bibinfo{year}{2013}).
\bibinfo{title}{{Three-nucleon forces in the 1/Nc expansion}}.
\bibinfo{journal}{{\em Phys. Rev. C}} \bibinfo{volume}{88}
  (\bibinfo{number}{3}): \bibinfo{pages}{034002}.
  \bibinfo{doi}{\doi{10.1103/PhysRevC.88.034002}}.
\eprint{1307.6274}.

\bibtype{Article}%
\bibitem[Ramsey-Musolf and Page(2006)]{Ramsey-Musolf:2006vfz}
\bibinfo{author}{Ramsey-Musolf MJ} and  \bibinfo{author}{Page SA}
  (\bibinfo{year}{2006}).
\bibinfo{title}{{Hadronic parity violation: A new view through the looking
  glass}}.
\bibinfo{journal}{{\em Ann. Rev. Nucl. Part. Sci.}} \bibinfo{volume}{56}:
  \bibinfo{pages}{1--52}.
  \bibinfo{doi}{\doi{10.1146/annurev.nucl.54.070103.181255}}.
\eprint{hep-ph/0601127}.

\bibtype{Article}%
\bibitem[Richardson and Schindler(2020)]{Richardson:2020iqi}
\bibinfo{author}{Richardson TR} and  \bibinfo{author}{Schindler MR}
  (\bibinfo{year}{2020}).
\bibinfo{title}{{Large-$N_c$ analysis of magnetic and axial two-nucleon
  currents in pionless effective field theory}}.
\bibinfo{journal}{{\em Phys. Rev. C}} \bibinfo{volume}{101}
  (\bibinfo{number}{5}): \bibinfo{pages}{055505}.
  \bibinfo{doi}{\doi{10.1103/PhysRevC.101.055505}}.
\eprint{2002.00986}.

\bibtype{Article}%
\bibitem[Richardson et al.(2023)]{Richardson:2022hyj}
\bibinfo{author}{Richardson TR}, \bibinfo{author}{Schindler MR} and
  \bibinfo{author}{Springer RP} (\bibinfo{year}{2023}).
\bibinfo{title}{{Implications of Large-Nc QCD for the NN Interaction}}.
\bibinfo{journal}{{\em Ann. Rev. Nucl. Part. Sci.}} \bibinfo{volume}{73}:
  \bibinfo{pages}{123--152}.
  \bibinfo{doi}{\doi{10.1146/annurev-nucl-102020-014052}}.
\eprint{2212.13049}.

\bibtype{Article}%
\bibitem[Sakharov(1967)]{Sakharov:1967dj}
\bibinfo{author}{Sakharov AD} (\bibinfo{year}{1967}).
\bibinfo{title}{{Violation of CP Invariance, C asymmetry, and baryon asymmetry
  of the universe}}.
\bibinfo{journal}{{\em Pisma Zh. Eksp. Teor. Fiz.}} \bibinfo{volume}{5}:
  \bibinfo{pages}{32--35}.
  \bibinfo{doi}{\doi{10.1070/PU1991v034n05ABEH002497}}.

\bibtype{Book}%
\bibitem[Scherer and Schindler(2012)]{Scherer:2012xha}
\bibinfo{author}{Scherer S} and  \bibinfo{author}{Schindler MR}
  (\bibinfo{year}{2012}).
\bibinfo{title}{{A Primer for Chiral Perturbation Theory}},
  \bibinfo{series}{Lect. Notes Phys.}, \bibinfo{volume}{830},
  \bibinfo{publisher}{Springer}.
\bibinfo{comment}{ISBN} \bibinfo{isbn}{978-3-642-19253-1}.
\bibinfo{doi}{\doi{10.1007/978-3-642-19254-8}}.

\bibtype{Article}%
\bibitem[Schindler and Springer(2013)]{Schindler:2013yua}
\bibinfo{author}{Schindler MR} and  \bibinfo{author}{Springer RP}
  (\bibinfo{year}{2013}).
\bibinfo{title}{{The Theory of Parity Violation in Few-Nucleon Systems}}.
\bibinfo{journal}{{\em Prog. Part. Nucl. Phys.}} \bibinfo{volume}{72}:
  \bibinfo{pages}{1--43}. \bibinfo{doi}{\doi{10.1016/j.ppnp.2013.05.002}}.
\eprint{1305.4190}.

\bibtype{Article}%
\bibitem[Schindler et al.(2018)]{Schindler:2018irz}
\bibinfo{author}{Schindler MR}, \bibinfo{author}{Singh H} and
  \bibinfo{author}{Springer RP} (\bibinfo{year}{2018}).
\bibinfo{title}{{Large-$N_c$ Relationships Among Two-Derivative Pionless
  Effective Field Theory Couplings}}.
\bibinfo{journal}{{\em Phys. Rev. C}} \bibinfo{volume}{98}
  (\bibinfo{number}{4}): \bibinfo{pages}{044001}.
  \bibinfo{doi}{\doi{10.1103/PhysRevC.98.044001}}.
\eprint{1805.06056}.

\bibtype{Book}%
\bibitem[Streater and Wightman(2001)]{Streater:1989vi}
\bibinfo{author}{Streater RF} and  \bibinfo{author}{Wightman AS}
  (\bibinfo{year}{2001}).
\bibinfo{title}{{PCT, spin and statistics, and all that}},
  \bibinfo{publisher}{Princeton}.
\bibinfo{comment}{ISBN} \bibinfo{isbn}{9780691070629}.

\bibtype{Article}%
\bibitem['t~Hooft(1973)]{tHooft:1973mfk}
\bibinfo{author}{'t~Hooft G} (\bibinfo{year}{1973}).
\bibinfo{title}{{Dimensional regularization and the renormalization group}}.
\bibinfo{journal}{{\em Nucl. Phys. B}} \bibinfo{volume}{61}:
  \bibinfo{pages}{455--468}. \bibinfo{doi}{\doi{10.1016/0550-3213(73)90376-3}}.

\bibtype{Article}%
\bibitem['t~Hooft(1974)]{tHooft:1973alw}
\bibinfo{author}{'t~Hooft G} (\bibinfo{year}{1974}).
\bibinfo{title}{{A Planar Diagram Theory for Strong Interactions}}.
\bibinfo{journal}{{\em Nucl. Phys. B}} \bibinfo{volume}{72}:
  \bibinfo{pages}{461}. \bibinfo{doi}{\doi{10.1016/0550-3213(74)90154-0}}.

\bibtype{Article}%
\bibitem['t~Hooft and Veltman(1972)]{tHooft:1972tcz}
\bibinfo{author}{'t~Hooft G} and  \bibinfo{author}{Veltman MJG}
  (\bibinfo{year}{1972}).
\bibinfo{title}{{Regularization and Renormalization of Gauge Fields}}.
\bibinfo{journal}{{\em Nucl. Phys. B}} \bibinfo{volume}{44}:
  \bibinfo{pages}{189--213}. \bibinfo{doi}{\doi{10.1016/0550-3213(72)90279-9}}.

\bibtype{Article}%
\bibitem['t~Hooft and Veltman(1979)]{tHooft:1978jhc}
\bibinfo{author}{'t~Hooft G} and  \bibinfo{author}{Veltman MJG}
  (\bibinfo{year}{1979}).
\bibinfo{title}{{Scalar One Loop Integrals}}.
\bibinfo{journal}{{\em Nucl. Phys. B}} \bibinfo{volume}{153}:
  \bibinfo{pages}{365--401}. \bibinfo{doi}{\doi{10.1016/0550-3213(79)90605-9}}.

\bibtype{Article}%
\bibitem[Vanasse and Phillips(2017)]{Vanasse:2016umz}
\bibinfo{author}{Vanasse J} and  \bibinfo{author}{Phillips DR}
  (\bibinfo{year}{2017}).
\bibinfo{title}{{Three-nucleon bound states and the Wigner-SU(4) limit}}.
\bibinfo{journal}{{\em Few Body Syst.}} \bibinfo{volume}{58}
  (\bibinfo{number}{2}): \bibinfo{pages}{26}.
  \bibinfo{doi}{\doi{10.1007/s00601-016-1173-2}}.
\eprint{1607.08585}.

\bibtype{Article}%
\bibitem[Vogel and Ormand(1993)]{Vogel:1993zz}
\bibinfo{author}{Vogel P} and  \bibinfo{author}{Ormand WE}
  (\bibinfo{year}{1993}).
\bibinfo{title}{{Spin-isospin SU(4) symmetry in sd- and fp-shell nuclei}}.
\bibinfo{journal}{{\em Phys. Rev. C}} \bibinfo{volume}{47}:
  \bibinfo{pages}{623--628}. \bibinfo{doi}{\doi{10.1103/PhysRevC.47.623}}.

\bibtype{Article}%
\bibitem[Weinberg(1966)]{Weinberg:1966kf}
\bibinfo{author}{Weinberg S} (\bibinfo{year}{1966}).
\bibinfo{title}{{Pion scattering lengths}}.
\bibinfo{journal}{{\em Phys. Rev. Lett.}} \bibinfo{volume}{17}:
  \bibinfo{pages}{616--621}. \bibinfo{doi}{\doi{10.1103/PhysRevLett.17.616}}.

\bibtype{Article}%
\bibitem[Weinberg(1968)]{Weinberg:1968de}
\bibinfo{author}{Weinberg S} (\bibinfo{year}{1968}).
\bibinfo{title}{{Nonlinear realizations of chiral symmetry}}.
\bibinfo{journal}{{\em Phys. Rev.}} \bibinfo{volume}{166}:
  \bibinfo{pages}{1568--1577}. \bibinfo{doi}{\doi{10.1103/PhysRev.166.1568}}.

\bibtype{Article}%
\bibitem[Weinberg(1973)]{Weinberg:1973un}
\bibinfo{author}{Weinberg S} (\bibinfo{year}{1973}).
\bibinfo{title}{{Nonabelian Gauge Theories of the Strong Interactions}}.
\bibinfo{journal}{{\em Phys. Rev. Lett.}} \bibinfo{volume}{31}:
  \bibinfo{pages}{494--497}. \bibinfo{doi}{\doi{10.1103/PhysRevLett.31.494}}.

\bibtype{Article}%
\bibitem[Weinberg(1979)]{Weinberg:1978kz}
\bibinfo{author}{Weinberg S} (\bibinfo{year}{1979}).
\bibinfo{title}{{Phenomenological Lagrangians}}.
\bibinfo{journal}{{\em Physica A}} \bibinfo{volume}{96}
  (\bibinfo{number}{1-2}): \bibinfo{pages}{327--340}.
  \bibinfo{doi}{\doi{10.1016/0378-4371(79)90223-1}}.

\bibtype{Article}%
\bibitem[Wigner(1937)]{Wigner:1936dx}
\bibinfo{author}{Wigner E} (\bibinfo{year}{1937}).
\bibinfo{title}{{On the Consequences of the Symmetry of the Nuclear Hamiltonian
  on the Spectroscopy of Nuclei}}.
\bibinfo{journal}{{\em Phys. Rev.}} \bibinfo{volume}{51}:
  \bibinfo{pages}{106--119}. \bibinfo{doi}{\doi{10.1103/PhysRev.51.106}}.

\bibtype{Article}%
\bibitem[Witten(1979)]{Witten:1979kh}
\bibinfo{author}{Witten E} (\bibinfo{year}{1979}).
\bibinfo{title}{{Baryons in the 1/N Expansion}}.
\bibinfo{journal}{{\em Nucl. Phys. B}} \bibinfo{volume}{160}:
  \bibinfo{pages}{57--115}. \bibinfo{doi}{\doi{10.1016/0550-3213(79)90232-3}}.

\end{thebibliography*}

\end{document}